\documentclass[trackchanges,twocolumn]{aastex701}

\renewcommand{\baselinestretch}{1.6}
\usepackage{tabularx}  
\usepackage{array}     
\usepackage{makecell}
\usepackage{amsmath} 
\usepackage{comment}
\usepackage{url}

\shortauthors{A. Gupta et al.}

\begin{document}

\title{Evidence for Solar-Cycle Modulation of the Alpha-to-Proton Temperature Ratio in Solar Wind}

\author[orcid=0009-0000-9917-2694,sname='Gupta']{Aakash Gupta}
\altaffiliation{Physical Research Laboratory (PRL), Ahmedabad-380009, India}
\affiliation{Physical Research Laboratory (PRL), Ahmedabad-380009, India}
\affiliation{Indian Institute of Technology (IIT), Gandhinagar-382055, India}
\email[show]{aakashgupta.du@gmail.com}  

\author[orcid=0000-0001-6018-9018, sname='Yogesh']{Yogesh} 
\affiliation{Department of Physics and Astronomy, University of Iowa, Iowa City IA 52242, USA}
\affiliation{NASA Goddard Space Flight Center, Greenbelt, MD, 20771, USA}
\affiliation{The Catholic University of America, Washington, DC 20064, USA}
\email{yphy22@gmail.com}

\author[orcid=0000-0003-2693-5325, sname='Chakrabarty']{Dibyendu Chakrabarty} 
\affiliation{Physical Research Laboratory (PRL), Ahmedabad-380009, India}
\email{dipu@prl.res.in}

\author[orcid=0000-0003-0602-6693, sname='Ofman']{Leon Ofman}
\affiliation{NASA Goddard Space Flight Center, Greenbelt, MD, 20771, USA}
\affiliation{The Catholic University of America, Washington, DC 20064, USA}
\affiliation{Visiting, Department of Geosciences, Tel Aviv University, Tel Aviv, Israel}
\email{ofman@cua.edu}

\author[orcid=0000-0003-1749-2665]{Gregory G. Howes}
\affiliation{Department of Physics and Astronomy, University of Iowa, Iowa City IA 52242, USA}
\email{ghowes@uiowa.edu}

\begin{abstract}
The influence of collisional age $(A_c)$ on the alpha-to-proton temperature ratio $(T_\alpha/T_p)$ has been explored in the past. However, the modulation of this ratio with respect to the solar cycle has remained unexplored so far. We show solar-cycle modulation of $T_\alpha/T_p$ and $A_c$ using nearly three decades of in-situ observations from Wind spacecraft across distinct solar wind speed regimes and solar activity phases. Our results reveal that in the slow solar wind with velocity $<400$ km s$^{-1}$, where $A_c$ happens to be typically $>1$, the ratio $T_\alpha/T_p$ stays close to unity. This suggests frequent Coulomb collisions efficiently iron out temperature differences. In contrast, the fast wind with velocity $>500$~km~s$^{-1}$, where $A_c$ happens to be typically $<1$, mass-proportional heating is most pronounced, with $T_\alpha/T_p$ often exceeding 4. The intermediate speed regime ($400$--$500$~km~s$^{-1}$) represents a gradual transition between the slow and fast wind populations in terms of their solar-cycle dependence. This behavior reflects the changing dominance of high-speed streams from polar coronal holes during minima to denser slow wind during maxima. These results suggest that  mass-proportional ion heating at 1 AU is not solely governed by local collisional physics but is significantly modulated by the solar cycle dependent variations in the solar wind sources.
\end{abstract}

\keywords{\uat{Solar wind}{1534} ---\uat{Sunspots}{1653} --- \uat{Solar coronal holes}{1484} --- \uat{Solar cycle}{1487} ---\uat{Heliosphere}{711}}

\section{Introduction} 

Solar wind is the hot tenuous plasma continuously flowing supersonically outward from the Sun’s outer atmosphere, the corona \citep{parker1958}. The bulk of the solar wind is composed mainly of protons ($\mathrm{H}^{+}$),  alpha ($\mathrm{He}^{++}$, $\sim 4\%$) particles, and electrons, with a small fraction contributed by heavier ions \citep{vonSteiger2000, Marsch2006}. The relative abundance of alpha particles has been studied in detail under different heliospheric conditions (i.e., fast and slow solar wind, interplanetary coronal mass ejections (ICMEs), and corotating-interaction regions (CIRs), etc.) using data and models \citep[e.g.,][and references therein]{Ogilvie1974, Aellig2001, Ofman2004, Kasper2007, Zerbo2015, Alterman2019, Yogesh2021, Alterman2021, Fu2018, Yogesh2022, Durovcova2019, Yogesh2023,Ofman2024, Yogesh2024}. While protons dominate in number density \citep{Bame1972}, the different plasma species exhibit distinct bulk velocities, differential streaming, temperature anisotropies, and beam populations, reflecting the weakly collisional and kinetically evolving nature of the solar wind \citep{Marsch1982, Marsch2012, Alterman2018, Durovcova2019, Verscharen2019, Verniero2022}. Observations from Helios, Ulysses, Wind and Parker Solar Probe (PSP)  have consistently revealed that alpha particles undergo preferential heating and acceleration compared to protons \citep{Ryan1975, Marsch1982, vonSteiger1995, Durovcova2019, Mostafavi2022, Mostafavi2025, Jagarlamudi2025}, highlighting a long-standing puzzle in heliophysics. 

In the context of solar-wind studies, it is important to note that the measured ion temperature does not represent heating alone. The temperature inferred from speed distribution functions reflects the combined effects of several physical processes: genuine plasma heating, adiabatic cooling associated with the radial expansion of the solar wind, collisional relaxation (including pitch-angle scattering), and the contribution of heat flux, which is typically small for ions heavier than protons.  A detailed discussion of these processes is provided in \citet{Cranmer2009}. In this study, therefore, we interpret variations in the alpha-to-proton temperature ratio ($T_{\alpha}/T_{p}$) as signatures of the net thermal evolution of the plasma, rather than heating alone.

The transition from the Sun’s visible photosphere with temperature of ~6000 K to its hot (million degree K) corona involves a steep rise in temperature within a narrow region of solar atmosphere with thickness of only a few hundred kilometers. In this transition region,  between ~0.1–0.3 $R_s$, the coronal plasma becomes effectively collisionless as the Coulomb collision frequency ($\nu_c \propto n/T^{3/2}$) drops sharply. In this environment, electrons and ions no longer remain in local thermodynamic equilibrium: ions become significantly hotter than electrons, and heavier ions attain higher temperatures than protons \citep{Landi2009}. Although such preferential ion heating may occur lower in the atmosphere, the higher collision rates at those altitudes would erase its signatures. Indeed, statistical studies reveal a broad range of relative ion temperatures, from equal values to mass-proportional and even super-mass-proportional scaling, with recent work suggesting approximate scaling as $T_i/T_p\sim (m_i/m_p)^{0.43}$ \citep{Tracy2016}. To explain these observations, several mechanisms of mass-proportional ion heating have been proposed. These include resonant absorption of ion-cyclotron waves \citep{Isenberg2009,Ofman2010,Jian2010,Maneva2013,Kasper2013, Navarro2020,Ofman2025, Yogesh2025}, interactions with low-frequency Alfv\'{e}nic turbulence \citep{Chandran2010, Chandran2013}, drift instabilities \citep{Verscharen2013, Martinovic2025}, impulsive reconnection events \citep{Cargill2004, Drake2009, Artemyev2014, Duan2023}, speed filtration \citep{Scudder1992}, and stochastic heating \citep{Chandran2010, Chandran2010b, Martinovic2020, Bowen2025}. Recent Parker Solar Probe observations further show strong radial evolution of proton temperature anisotropy, plasma beta, Alfv\'enic fluctuations, and magnetic-field turbulence across sub-Alfv\'enic and super-Alfv\'enic regions, suggesting that wave-particle interactions and turbulent fluctuations provide an important source of free energy for particle heating in the near-Sun solar wind \citep[and references within]{Yogesh2026}. These observations are supported by linear Vlasov Maxwell calculations showing that oblique drift instabilities driven by proton beam or alpha particle populations can reduce differential streaming and heat ions, thereby contributing to preferential ion heating in the low ($\beta$), near-Sun solar wind \citep{Mihailo2026}.

Collisionality provides a framework for characterizing how these processes manifest in the expanding solar wind. The collisional age ($A_c$) \citep{Kasper2008, Kasper2017} represents the cumulative number of Coulomb collisions experienced by the plasma during its expansion. Analyses of Wind observations at 1 AU showed that solar wind properties are strongly organized by $A_c$ \citep{Kasper2008, Maruca2013, Alterman2018}. Since protons are the most abundant ion species in the solar wind, the proton collisional age is given by

\begin{equation}
\begin{aligned}
A_c ={}&
\left( 1.31 \times 10^{7}\,
\frac{\mathrm{cm}^3\,\mathrm{km}\,\mathrm{K}^{3/2}}
{\mathrm{s}\,\mathrm{AU}} \right)
\left( \frac{n_p}{v_{sw}\,T_p^{3/2}} \right) r \\
&\times
\left[
9.42 + \ln \left(
\frac{1}{\mathrm{cm}^{3/2}\,\mathrm{K}^{3/2}}
\left( \frac{T_p^{3/2}}{n_p^{1/2}} \right)
\right)
\right]
\end{aligned}
\label{eq:collisional_age}
\end{equation}


Here, $n_p$ denotes the proton number density, $v_{sw}$ is the radial solar wind speed, $T_p$ is the proton temperature, and $r$ is the heliocentric distance of the spacecraft, which in this study is $1\,\mathrm{AU}$. When $A_c < 1$, the plasma is effectively collisionless, and nonthermal features, such as alpha-proton temperature disequilibrium and differential streaming, are commonly observed because Coulomb collisions are insufficient to efficiently relax these kinetic signatures. Conversely, when $A_c > 1$, the plasma is "collisionally old," and Coulomb collisions act to reduce or erase these nonthermal signatures\citep{Maruca2013, Kasper2017, Alterman2018, Kasper2019,  Martinovic2020, Johnson2023, mostafavi_et_al_2024, Martinovic2025}. 

This transition from collisionless to collisional regimes is highlighted by recent observations from Parker Solar Probe (PSP), which have extended the investigation of mass-proportional ion heating and collisional evolution into the near-Sun environment. In these regions, the plasma is significantly less collisionally processed than at $1\,\mathrm{AU}$. PSP measurements have revealed strong radial evolution of proton and alpha particle temperatures, temperature anisotropies, and differential streaming under conditions of extremely low Coulomb number, providing new constraints on the role of collisional relaxation versus wave-particle interactions in shaping ion thermodynamics \citep{Ofman2022, Ofman2023, mostafavi_et_al_2024, mario_et_al_2024, peng_et_al_2024, Ofman2025}. These near-Sun observations offer an important comparison point to $1\,\mathrm{AU}$ studies, where the cumulative effects of collisional processing, quantified through parameters such as $A_c$, are more pronounced. Consistent with this picture, Wind data show that the alpha-to-proton temperature ratio ($T_\alpha/T_p$) and the field-aligned alpha-proton differential flow,
$(\mathbf{V}_{\alpha}-\mathbf{V}_{p})\cdot\hat{\mathbf{b}}$,
where $\mathbf{V}_{\alpha}$ and $\mathbf{V}_{p}$ are the alpha-particle and proton bulk velocity vectors, respectively, and $\hat{\mathbf{b}}$ is the unit vector along the magnetic field, correlate positively with solar wind speed \citep{Kasper2008, Kasper2017, Maruca2013, Alterman2018}. Thus, slow solar wind exhibits large collisional ages wherein collisions act to regulate nonthermal effects in protons and alpha particles and can drive the plasma toward thermal equilibrium \citep{Feldman1974, Neugebauer1976, Livi1986, Marsch1983, Hernandez1985, Klein1985, Kasper2008, Tracy2015, Alterman2018, Durovcova2021}. 

In contrast, fast solar wind tends to preserve non-equilibrium states with enhanced temperature ratios and differential streaming under weakly collisional conditions. Recent observations from the Aditya-L1 mission have further expanded our understanding of solar wind ion populations under both quiet and disturbed heliospheric conditions. Using measurements from the Aditya Solar Wind Particle Experiment \citep[ASPEX;][]{Kumar2025, Goyal2025, Sebastian2025, Sebastian2026}, recent studies have characterized the properties of quiet-time energetic ions \citep{Gupta2025} and reported proton and alpha-particle energization within an ICME-ICME interaction region \citep{Parashar2026}, highlighting the role of large-scale heliospheric structures, shocks, and interaction regions in shaping the thermodynamic properties of solar wind ions during interplanetary propagation. Understanding how these collisional and collisionless processes operate in the solar wind remains one of the key open challenges. Although several studies have investigated the preferential or mass-proportional heating of ions (protons and alpha particles) in the solar wind \citep[e.g.,][]{Kasper2013, Kasper2017, Kasper2019, Maruca2013}, the underlying physical mechanisms responsible for this behavior remain an active topic of research. In particular, the dependence of the temperature ratio $T_\alpha/T_p$ on collisional age ($A_c$) across distinct solar wind speed regimes and over different phases of the solar cycle has not yet been comprehensively characterized. 

The present investigation addresses this gap by examining the combined effects of solar wind speed, collisional processing, and solar cycle evolution on mass-proportional ion heating. The structure of the paper is as follows: Section 2 describes the dataset used in this study, Section 3 presents the data analysis and key observational results, Section 4 provides the discussion, and section 5 presents the conclusion drawn from the study.

\section{Dataset} \label{sec:Dataset}

In this study, we have utilized a large dataset of bulk parameters derived from the Solar Wind Experiment (SWE; \citet{Ogilvie1995}) on-board the NASA Wind spacecraft \citep{Acuna1995}, covering the period from 1 January 1995 to 31 December 2024. The SWE Faraday Cup instruments on-board Wind measure solar wind $\mathrm{H}^{+}$ (protons) and $\mathrm{He}^{++}$ (alpha-particles), recording detailed three-dimensional velocity distribution functions (VDFs) of these ion species at a cadence of $\sim 92$ seconds \citep{Kasper2007}. From this dataset, the final selection of data is made based on the following four criteria: (1) The Wind spacecraft had to be well outside the Earth’s bow shock, i.e., in the undisturbed solar wind medium. We did not check for ion or electron foreshock regions, so no foreshock exclusion was applied. The corresponding dates when Wind was inside the bow shock are obtained from the NASA Wind bow shock database \footnote{\url{https://wind.nasa.gov/mfi/bow_shock.html}}, and those intervals are filtered out from the analysis. This step is important since, during the early phase of the mission, the Wind spacecraft spent a considerable amount of time exploring Earth’s magnetosphere. (2) All interplanetary coronal mass ejection (ICME) intervals were excluded from the dataset using the Wind ICME catalog \footnote{\url{https://wind.nasa.gov/ICME_catalog/ICME_catalog_viewer.php}} to ensure that only ambient solar wind conditions were analyzed. (3) The ion VDF fit results are required to be of high quality, for which we used only intervals where the ion VDF fit flag = 10. (4) Only those ion bulk parameters derived from the VDF fittings were considered in the analysis when the corresponding fitting uncertainties were less than 10\% of the fitted parameter values. For studying solar cycle variations, the monthly mean sunspot number dataset from the WDC-SILSO, Royal Observatory of Belgium, Brussels, is also used \citep{SILSO_Sunspot_Number}. The dataset is publicly available at \href{https://doi.org/10.24414/qnza-ac80}{SILSO data, 2024}.

\section{Data Analysis and Observations} \label{sec:Data Analysis and Observations}

\begin{figure*}[ht!]
\plotone{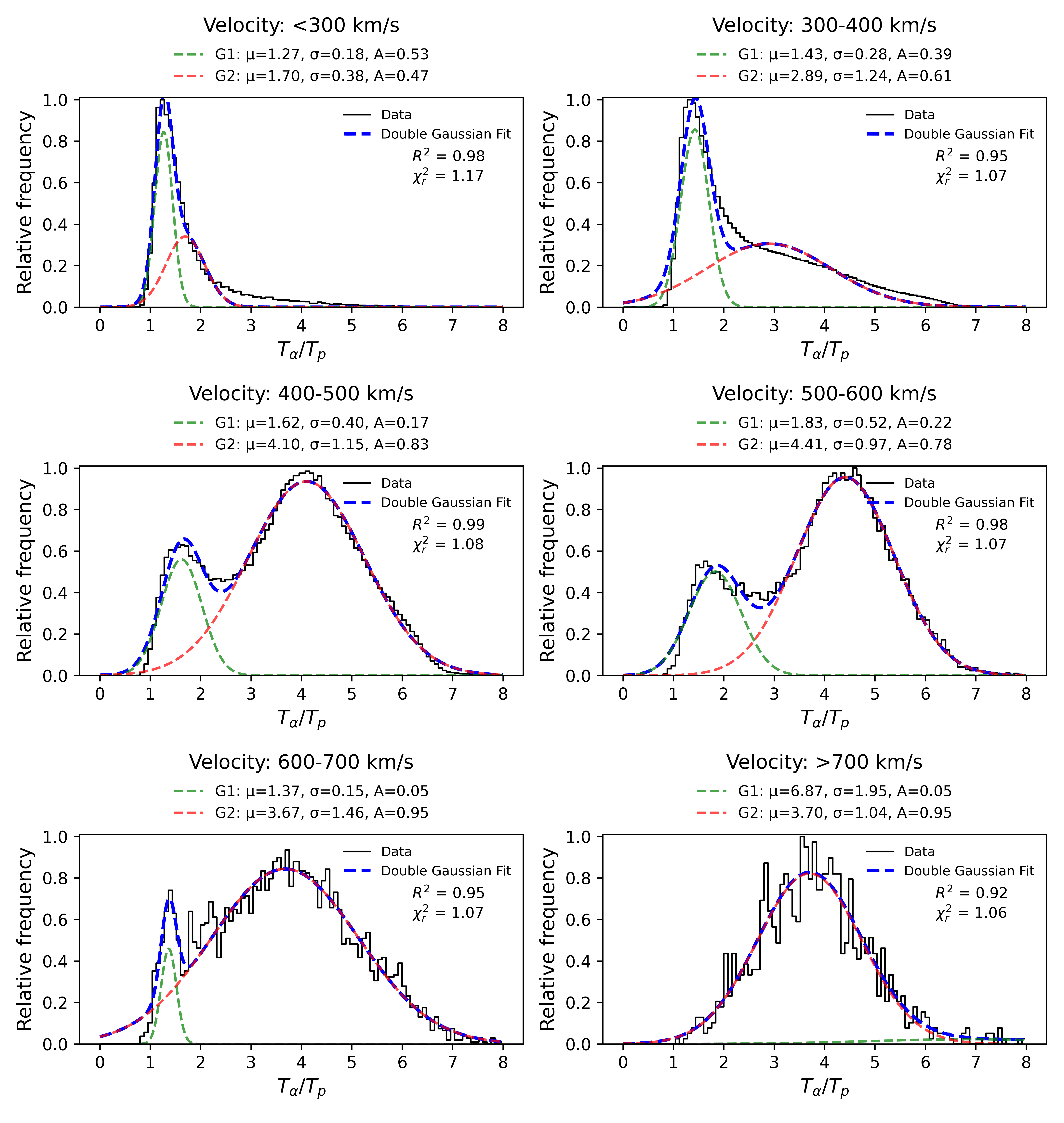}
\caption{Distribution of the alpha-to-proton temperature ratio ($T_\alpha/T_p$) in different solar wind speed bins, fitted with a double-Gaussian model. Each panel corresponds to the speed range indicated at the top: $<300$, $300$–$400$, $400$–$500$, $500$–$600$, $600$–$700$, and $>700$ km s$^{-1}$. The black step histograms represent the observed relative frequency of $T_\alpha/T_p$. The blue dashed curves show the total double-Gaussian fit to the data. The individual Gaussian components are overplotted as green dashed lines ($G_1$) and red dashed lines ($G_2$), corresponding respectively to the equal alpha-to-proton temperature ratio population and the mass-proportional temperature ratio population. The fitted parameters mean ($\mu$), standard deviation ($\sigma$), and fractional area ($A$) for both components are listed at the top of each panel. The goodness of fit is quantified by the coefficient of determination ($R^2$) and reduced chi-square ($\chi_r^2$), also indicated within each panel. At low velocities ($<400$ km s$^{-1}$), the $T_\alpha/T_p$ distribution is dominated by an equal alpha-to-proton temperature ratio component, whereas at higher velocities ($> 400$ km s$^{-1}$), a distinct mass-proportional temperature component emerges with broader distributions and higher $T_\alpha/T_p$ values ($\sim4$), indicative of stronger mass-proportional heating in the fast, weakly collisional solar wind.
\label{fig:general}}
\end{figure*}

The investigation of mass-proportional heating between protons and alpha particles, representing the non-equilibrium nature of the solar wind plasma, involves calculating $T_\alpha/T_p$. Since different ion species in the solar wind exhibit distinct bulk speed and anisotropic temperatures along directions parallel and perpendicular to the background magnetic field, the isotropic temperature of each species is evaluated using the relation $T_{j} = (2T_{\perp j} + T_{\parallel j}) / 3$, where j denotes the ion species. To investigate the variation of mass-proportional heating of alpha particles across different solar wind speed ranges, the calculated $T_\alpha/T_p$ are categorized into six speed bins: $<300$ km s$^{-1}$, $300–400$ km s$^{-1}$, $400–500$ km s$^{-1}$, $500–600$ km s$^{-1}$, $600–700$ km s$^{-1}$, and $>700$ km s$^{-1}$. Solar wind bulk speed is commonly used as a proxy for distinguishing plasma originating from different coronal source regions and subjected to different heating and acceleration histories \citep{McIntosh2011}. Previous studies have employed a variety of classification schemes, including broad slow and fast wind intervals \citep{McIntosh2011, Alterman2025a}, fixed speed bins \citep{Kasper2012}, and speed quantiles \citep{Alterman2019, Alterman2021}. In this work, we adopt fixed 100 km s$^{-1}$ speed intervals because our objective is to investigate how the distribution of $T_\alpha/T_p$ evolves with the absolute solar wind speed. Although quantile-based binning provides comparable sample sizes across bins, fixed speed intervals preserve the physical speed scale and enable direct comparisons between distinct solar wind regimes. The chosen bin width provides sufficient statistical sampling while retaining sensitivity to variations in the relative contributions of the two commonly observed populations near $T_\alpha/T_p \sim 1$ and $T_\alpha/T_p \sim 4$ across the solar wind speed range. The relative frequency distributions of $T_\alpha/T_p$ within each speed interval are shown in Figure 1.

In Figure 1, the black histograms show the observed data, while the blue dashed curves represent the total double Gaussian fit to the distribution. The two Gaussian components are plotted separately, with the green dashed line ($G_1$) and the red dashed line ($G_2$) representing the Gaussian fittings for an equal alpha-to-proton temperature ratio and mass proportional temperature ratio. For each speed bin, the mean ($\mu$), width ($\sigma$), and fractional contribution (area under the curve, A) of both Gaussian components are provided at the top of each panel. It can be seen that at lower solar wind speeds ($<400$ km s$^{-1}$), the distributions are sharply peaked near $T_\alpha/T_p \sim 1$, dominated by equal alpha and proton temperatures of the solar wind, with only a minor contribution from the mass-proportional temperature ratio component. As the solar wind speed increases, the relative contribution of the mass-proportional temperature component becomes more significant. At higher velocities ($>400$ km s$^{-1}$), the distributions broaden and reveal two distinct populations, with $T_\alpha/T_p \sim 1$ and $T_\alpha/T_p  \sim 4$. The mass-proportional temperature component indicates that alphas and protons have equal thermal speeds instead of equal temperatures. In the highest speed regimes ($>600$ km s$^{-1}$), the mass-proportional temperature component (red Gaussian) becomes dominant, shifting the distribution towards $T_\alpha/T_p  \sim 4$. 


\subsection{Solar cycle dependence of mass-proportional heating in protons and alpha particles }  

\begin{figure*}[ht!]
\plotone{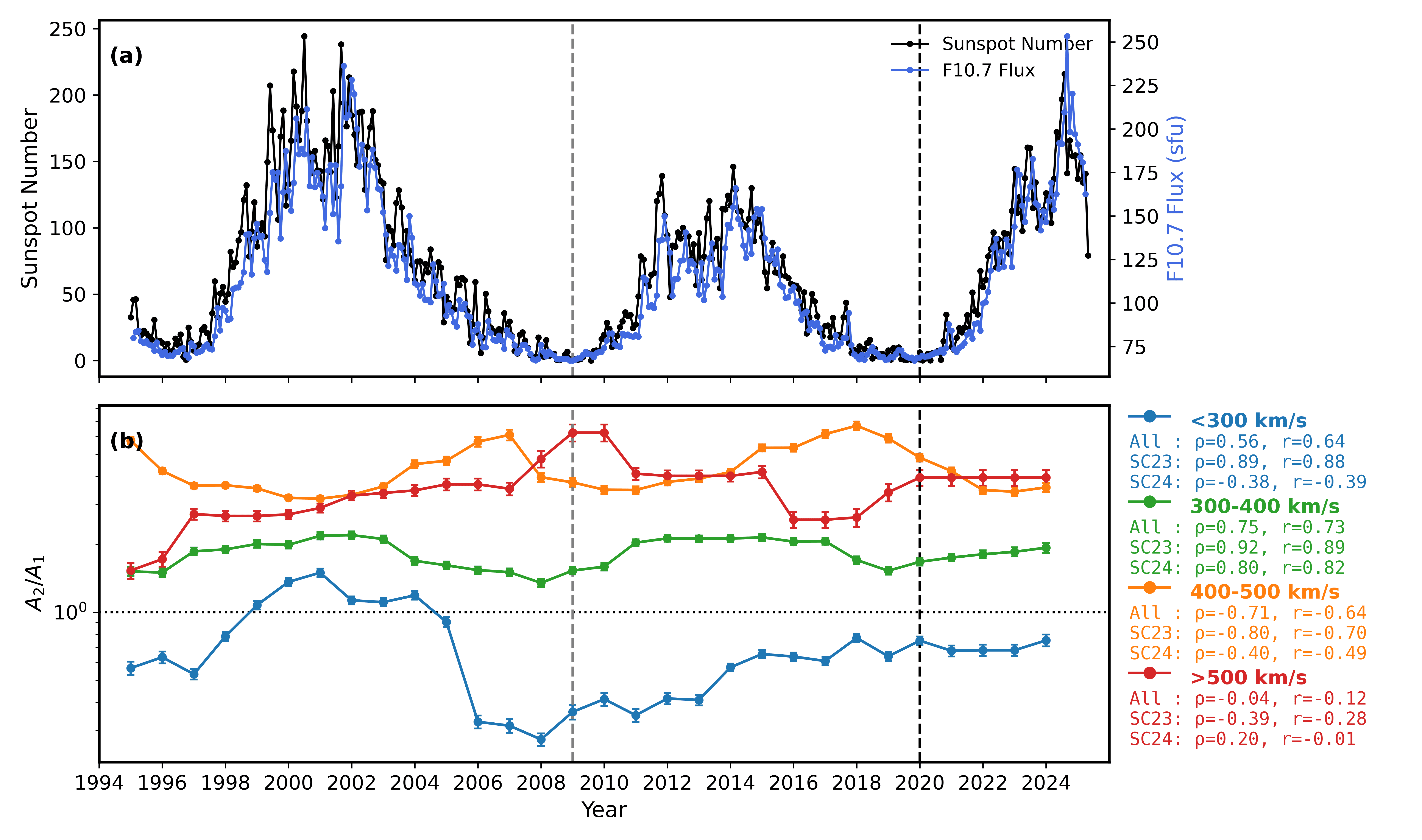}
\caption{Solar-cycle evolution of the ratio between the mass-proportional and equal-temperature populations, represented by the fitted double-Gaussian component area ratio \(A_2/A_1\), for different solar wind speed regimes. Panel (a) shows the monthly averaged sunspot number (black) and F10.7 solar radio flux (blue) as proxies of solar activity. Panel (b) presents the solar cycle variation of \(A_2/A_1\) for four solar wind speed bins: \(<300\) km s\(^{-1}\) (blue), \(300\)-\(400\) km s\(^{-1}\) (green), \(400\)-\(500\) km s\(^{-1}\) (orange), and \(>500\) km s\(^{-1}\) (red). Here, \(A_1\) corresponds to the area of Gaussian component centered near \(T_\alpha/T_p \sim 1\), representing the equal-temperature population, while \(A_2\) corresponds to the area of Gaussian component centered near \(T_\alpha/T_p \sim 4\), representing the mass-proportional temperature population. Values of \(A_2/A_1 > 1\) indicate dominance of the mass-proportional population, whereas \(A_2/A_1 < 1\) indicate dominance of the equal-temperature population. Error bars denote the propagated \(1\sigma\) uncertainties obtained from the double-Gaussian fitting procedure. The Spearman rank correlation coefficients (\(\rho\)) and Pearson correlation coefficients (\(r\)) between \(A_2/A_1\) and solar activity are listed separately for Solar Cycle 23 (SC23), Solar Cycle 24 (SC24), and the full interval (“All”) for each speed bin. The horizontal dotted line marks \(A_2/A_1 = 1\), where the two populations contribute equally. The gray vertical dashed line marks the transition between SC23 and SC24, while the black vertical dashed line marks the transition between SC24 and SC25.}
\label{fig:general}
\end{figure*}

The solar cycle dependence of the relative occurrence of mass-proportional and equal-temperature plasma populations is examined through the temporal evolution of the ratio of the fitted double-Gaussian areas, $A_2/A_1$, where $A_1 = \int G_1(\tilde{T})  d\tilde{T}$)  represents the population centered near \(T_\alpha/T_p \sim 1\) and $A_2 = \int G_2(\tilde{T}) d\tilde{T}$ represents the population centered near \(T_\alpha/T_p \sim 4\). Here $\tilde{T}$ represents the ratio of $T_\alpha$ and $T_p$. Values of \(A_2/A_1 > 1\) indicate that the mass-proportional population dominates the distribution, whereas \(A_2/A_1 < 1\) corresponds to the dominance of the equal-temperature population. Figure 2 presents the solar cycle evolution of this ratio $(A_2/A_1)$ for four solar wind speed bins together with monthly sunspot number and F10.7 solar radio flux, which serve as independent proxies of solar activity. The error bars represent the propagated \(1\sigma\) uncertainties obtained from the double-Gaussian fitting.

To quantify the solar-cycle dependence, both the Spearman rank correlation coefficient (\(\rho\)) and the Pearson correlation coefficient (\(r\)) were calculated between \(A_2/A_1\) and solar activity. The correlations were evaluated for the complete interval (1995-2024) as well as separately for Solar Cycle 23 (1996-2008) and Solar Cycle 24 (2009-2020). The use of Spearman correlations allows identification of monotonic relationships (not necessarily linear), while the Pearson coefficient measures the corresponding linear dependence. The generally good agreement between \(\rho\) and \(r\) indicates that the observed trends are both monotonic and approximately linear over the solar cycle timescales considered here.

The strongest solar-cycle dependence is observed in the slow and intermediate-speed solar wind. In the \(<300\) km s\(^{-1}\) bin, the ratio \(A_2/A_1\) remains below unity throughout most of the interval, indicating that the equal-temperature population dominates the distribution. Nevertheless, the ratio exhibits a clear positive correlation with solar activity (\(\rho=0.56\), \(r=0.64\) for the full interval), implying that the relative contribution of the mass-proportional population increases during periods of enhanced solar activity. Similar behavior is observed in the \(300\)-\(400\) km s\(^{-1}\) range, where \(A_2/A_1\) consistently exceeds unity, indicating that the mass-proportional population starts dominating. This speed bin also exhibits the strongest positive solar-cycle dependence (\(\rho=0.75\), \(r=0.73\)). The correlations are particularly strong during Solar Cycle 23 and remain significant during Solar Cycle 24, indicating that the modulation persists across multiple solar cycles.

The \(400\)-\(500\) km s\(^{-1}\) speed range represents a transition regime. Here, the mass-proportional population remains dominant (\(A_2/A_1 > 1\)), but the ratio becomes negatively correlated with solar activity (\(\rho=-0.71\), \(r=-0.64\)). This reversal indicates that the relative abundance of plasma exhibiting mass-proportional temperatures increases during solar minima and decreases during solar maxima. 

In the fast solar wind (\(>500\) km s\(^{-1}\)), the ratio \(A_2/A_1\) remains substantially greater than unity throughout the entire interval, demonstrating the persistent dominance of the mass-proportional temperature population. Unlike the lower-speed bins, however, the solar-cycle dependence is weak (\(\rho=-0.04\), \(r=-0.12\) for the full interval), indicating that the fast wind temperature-ratio distribution is comparatively insensitive to changes in solar activity. Although moderate cycle-to-cycle differences are present, the consistently large values of \(A_2/A_1\) show that mass-proportional temperatures remain a characteristic feature of the fast solar wind regardless of solar-cycle phase.

\subsection{Distribution of $T_\alpha/T_p$ with collisional age ($A_c$) within distinct solar wind speed regimes}

\begin{figure*}[ht!]
\plotone{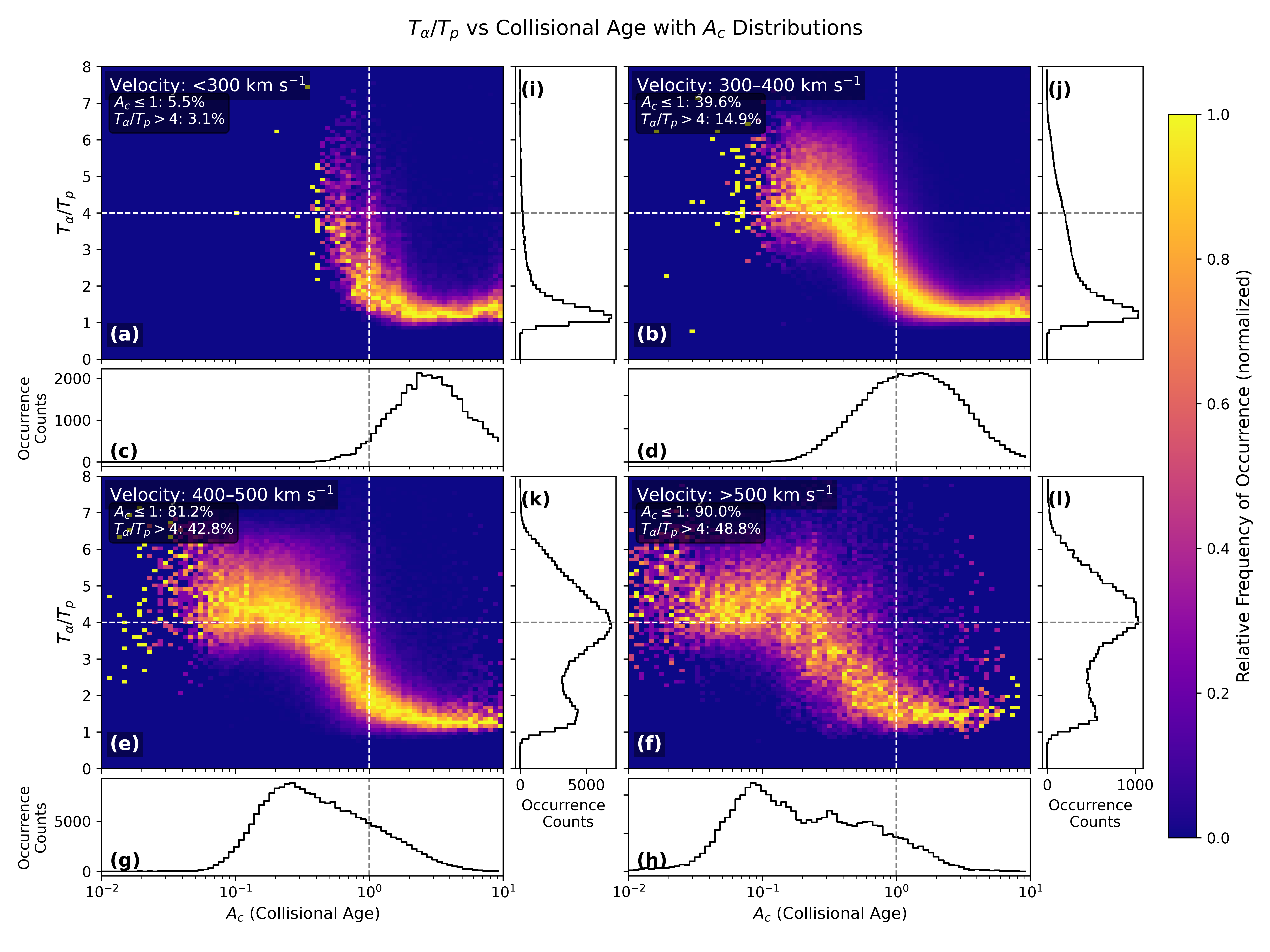}
\caption{This figure shows the variation of the temperature ratio $T_\alpha/T_p$ as a function of the Coulomb collisional age ($A_c$) for different solar wind speed bins: (a) $<300$ km s$^{-1}$, (b) $300$--$400$ km s$^{-1}$, (e) $400$--$500$ km s$^{-1}$, and (f) $>500$ km s$^{-1}$. The color scale represents the normalized relative frequency of occurrence. The dashed vertical line marks $A_c = 1$, distinguishing weakly ($A_c < 1$) and strongly ($A_c > 1$) collisional regimes, while the horizontal dashed line at $T_\alpha/T_p = 4$ indicates the threshold for significant mass-proportional heating of alpha particles. The inset boxes in each panel list the percentage of data points within the weakly collisional regime ($A_c < 1$) and those exhibiting enhanced heating ($T_\alpha/T_p > 4$. Panels (c), (d), (g), and (h) show the corresponding distributions of measurement occurrences as a function of collisional age for $<300$ km s$^{-1}$, $300$--$400$ km s$^{-1}$, $400$--$500$ km s$^{-1}$, and $>500$ km s$^{-1}$ speed bin respectively, providing the relative population of plasma intervals contributing to the two-dimensional distributions above. Panels (i), (j), (k), and (l) show the corresponding marginal distributions of $T_\alpha/T_p$ for the same speed bins, illustrating the systematic shift toward higher temperature ratios with increasing solar wind speed. The results demonstrate that mass-proportional heating is predominantly observed in the weakly collisional, high-speed solar wind, where collisional relaxation is minimal.}
\label{fig:general} 
\end{figure*}

Figure 3 presents the distribution of $T_\alpha/T_p$ as a function of collisional age ($A_c$) for four distinct solar wind speed bins. By examining the distributions within relatively narrow velocity ranges, the analysis minimizes the direct influence of the explicit inverse dependence of $A_c$ on solar wind speed in Equation 1 and allows the variation of $T_\alpha/T_p$ with collisionality to be investigated within approximately fixed solar wind regimes. Figure 3 clearly illustrates the interplay between solar wind speed, collisionality, and the mass-proportional heating of alpha particles. In the slow solar wind ($<300$ km s$^{-1}$), the majority of the distribution corresponds to $A_c > 1$ (marked by vertical white dashed line). Note, in the figures, $T_\alpha/T_p = 4$ is marked by horizontal white dashed line. In the solar wind speed ranges of $300$–$400$ km s$^{-1}$ and $400$–$500$ km s$^{-1}$, a clear transition occurs as solar wind speed increases: the typical collisional age within each bin shifts to lower values. Correspondingly, the distribution broadens and an increasing fraction of the plasma exhibits $T_\alpha/T_p > 4$, indicating a gradual reduction in collisional thermalization and a transition toward mass-proportional alpha-particle temperatures.

Figure 3 further shows that the fraction of intervals with $A_c < 1$ and $T_\alpha/T_p > 4$ systematically increases from the slow to the fast solar wind regimes. For example, the fraction of plasma with $A_c < 1$ increases from $\sim5.5\%$ in the $<300$ km s$^{-1}$ bin to $\sim90\%$ in the $>500$ km s$^{-1}$ bin, while the occurrence of $T_\alpha/T_p > 4$ simultaneously increases from $\sim3.1\%$ to $\sim48.8\%$. Correspondingly, the spread in $T_\alpha/T_p$ broadens significantly toward higher solar wind speeds. In the fastest solar wind ($>500$ km s$^{-1}$), the vast majority of the plasma lies firmly in the low-collisionality regime ($A_c < 1$), and the distribution exhibits a substantial population with $T_\alpha/T_p \gtrsim 4$. These observational trends indicate a strong association between low collisional age and the persistence of enhanced alpha-to-proton temperature ratios across different solar wind regimes \citep{Kasper2017, Kasper2019, Maruca2013, Johnson2023, Johnson2024, Johnson2025}.


\subsection{Solar cycle dependence of collisionality and mass-proportional heating}

Figure 4 presents the solar-cycle dependence of solar wind collisionality and the occurrence of enhanced alpha-to-proton temperature ratios across different solar wind speed bins. Panel (a) shows the monthly sunspot number and F10.7 solar radio flux, which are used as proxies for solar activity. Panel (b) shows the temporal evolution of the fraction of weakly collisional plasma ($A_c < 1$) for each speed bin. Panel (c) shows the temporal evolution of the fraction of intervals with ($T_\alpha/T_p > 4$), representing the occurrence of mass-proportional alpha-to-proton temperature ratios. For both panels (b) and (c), the Spearman rank correlation coefficients ($\rho$) and Pearson correlation coefficients ($r$) between the fractions namely,  \(A_c < 1\) in panel (b) and \(T_\alpha/T_p > 4\) in panel (c), and solar activity are indicated in the legends for each speed bin. Panel (d) summarizes these Spearman rank correlation coefficients ($\rho$) and Pearson correlation coefficients ($r$) as a function of solar-wind speed bin. Here, the threshold $A_c < 1$ is adopted to identify plasma that is effectively collisionless on the scale of solar wind expansion to 1 AU, corresponding to conditions where particles undergo, on average, fewer than one significant Coulomb collision during transit. This broader criterion is widely used in statistical solar wind studies to distinguish relatively collisionless (“young”) wind from collisionally processed (“old”) wind \citep{Maruca2013, Kasper2017}.

\begin{figure*}[ht!]
\includegraphics[height=0.78\textheight, width=\textwidth]{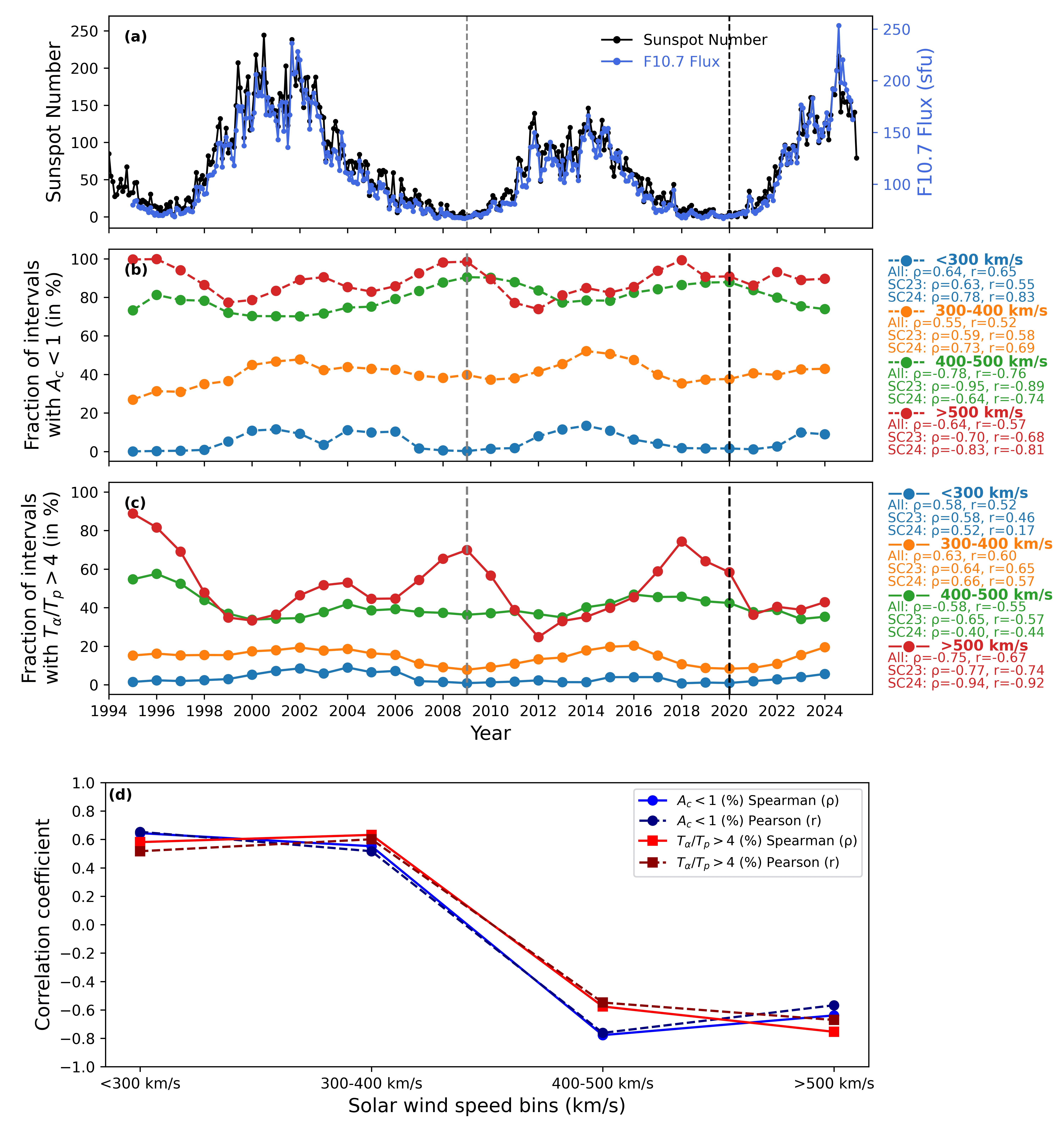}
\caption{Solar cycle modulation of solar wind collisionality and mass-proportional alpha particle heating across different wind speed regimes. The panel (a) presents the monthly sunspot number (black) and F10.7 solar radio flux (blue) as indicators of solar activity from 1995 to 2024. The panel (b) shows the percentage of collisionless solar wind intervals ($A_c < 1$) for various speed bins, with the corresponding Spearman rank correlation coefficients ($\rho$) and Pearson correlation coefficients ($r$) with solar activity separately for Solar Cycle 23 (SC23), Solar Cycle 24 (SC24), and the full interval (“All”) indicated in the legend. The panel (c) displays the percentage occurrence of strong alpha particle heating ($T_\alpha/T_p > 4$) in each speed bin, again with the corresponding Spearman rank correlation coefficients ($\rho$) and Pearson correlation coefficients ($r$) with solar activity separately for Solar Cycle 23 (SC23), Solar Cycle 24 (SC24), and the full interval (“All”) shown in the legend. The panel (d) summarizes the Spearman rank correlation coefficients ($\rho$) and Pearson correlation coefficients ($r$) between the occurrence percentages shown in panels (b) and (c) and solar activity as a function of solar wind speed bin. The gray vertical dashed line marks the transition between SC23 and SC24, while the black vertical dashed line marks the transition between SC24 and SC25.}
\label{fig:general}
\end{figure*} 

In Figure 4, panel (b) shows that in the lower solar wind speed bins ($<300$ and $300$–$400$ km s$^{-1}$), the percentage of $A_c < 1$ closely follows the solar cycle trend, consistent with the positive correlation values ($\rho = 0.64$, $r = 0.65$ and $\rho = 0.55$, $r = 0.52$, respectively), with higher values during solar maxima and lower values during minima. However, in the intermediate speed range ($400$–$500$ km s$^{-1}$), this behavior reverses: the fraction of $A_c < 1$ is slightly enhanced during solar minima and reduced during solar maxima, opposite to the trend observed at lower speeds, as reflected by the negative correlations ($\rho = -0.78$, $r = -0.76$). A similar anti-correlation is observed in the $>500$ km s$^{-1}$ bin ($\rho = -0.64$, $r = -0.57$), indicating that the fastest wind tends to be more collisionless during solar minima. In $>500$ km s$^{-1}$ speed bin, the solar wind is dominantly young and weakly collisional, with more than $80\%$ of the intervals characterized by $A_c < 1$. 

A similar reversal is evident in panel (c) for the fraction of $T_\alpha/T_p > 4$. Specifically, in the intermediate speed bin, the occurrence of $T_\alpha/T_p > 4$ is elevated during solar minima and suppressed during maxima, consistent with the negative correlations ($\rho = -0.58$, $r = -0.55$). In contrast, the lower speed bins show positive correlations ($\rho = 0.58$, $r = 0.52$ for $<300$ km s$^{-1}$ and $\rho = 0.63$, $r = 0.60$ for $300$–$400$ km s$^{-1}$), indicating enhanced high temperature ratios during solar maxima in slow wind. In the $>500$ km s$^{-1}$ speed bin, a strong anti-correlation ($\rho = -0.75$, $r = -0.67$) is observed, confirming that preferential alpha heating is most pronounced during solar minima in fast wind. Taken together, panel (d) of Figure 4 summarizes this speed-dependent reversal: the correlations of both the weakly collisional fraction ($A_c < 1$) and the fraction of intervals with $T_\alpha/T_p > 4$ are positive in the slow wind speed bins but become strongly negative in the intermediate and fast wind speed bins. The transition occurs between the $300–400$ and $400–500$ km s$^{-1}$ speed bins.

\section{Discussion}

This study provides observational evidence that the alpha-to-proton temperature ratio in the solar wind strongly depends on solar wind speed, collisional age, and solar-cycle phase. The distributions of $T_\alpha/T_p$ reveal two dominant populations corresponding to near-equal temperatures ($T_\alpha/T_p \sim 1$) and mass-proportional temperatures ($T_\alpha/T_p \sim 4$), whose relative occurrence changes systematically with solar wind speed. The transition from an equal-temperature population in the slow solar wind to a mass-proportional population in the fast solar wind highlights the changing nature of ion heating and energy partitioning across different solar wind regimes.

To assess the robustness of these results, and to account for the positive-definite and mildly skewed nature of the $T_\alpha/T_p$ distributions, we analyzed the data using both a double-Gaussian and a two-component lognormal mixture model. Both the double-Gaussian and double-lognormal representations successfully reproduce the observed bimodal distributions and recover the same speed-dependent and solar-cycle evolution of the relative population fractions, demonstrating that the physical conclusions are insensitive to the choice of mixture model. However, the lognormal representation yields systematically larger propagated $1\sigma$ uncertainties in the fitted component areas. This is because the two lognormal components overlap more strongly, making their relative contributions less well constrained and leading to larger uncertainties in the estimated population fractions. In contrast, the double-Gaussian model provides more stable parameter estimates and smaller uncertainties in the fractional areas, which are the primary quantities tracked throughout the solar-cycle analysis. For clarity and statistical stability, the double-Gaussian model was therefore adopted in the main analysis.

In the slow solar wind ($<300$ km s$^{-1}$ and $300-400$ km s$^{-1}$), where the collisional age is typically $A_c > 1$, the distributions are dominated by the population centered near $T_\alpha/T_p \sim 1$, indicating that intervals with nearly equal proton and alpha-particle temperatures occur most frequently. As the solar wind speed increases, the relative contribution of the higher-temperature population progressively increases. In the fastest solar wind ($>500$ km s$^{-1}$), most intervals satisfy $A_c < 1$, and the distributions are instead dominated by the population centered near $T_\alpha/T_p \sim 4$, with a substantial fraction of intervals exhibiting mass-proportional or even super-mass-proportional temperature ratios. The intermediate-speed solar wind ($400$-$500$ km s$^{-1}$) exhibits properties between these two limiting cases, reflecting a gradual transition rather than an abrupt change in plasma characteristics.

The close correspondence between solar wind speed, collisional age, and the relative occurrence of the two temperature populations is further reflected in the occurrence fractions of $A_c < 1$ and $T_\alpha/T_p > 4$. As shown in Figure~4, weakly collisional plasma ($A_c < 1$) and enhanced alpha-particle temperature ratios become increasingly common with increasing solar wind speed, while intervals characterized by $A_c > 1$ and $T_\alpha/T_p \sim 1$ dominate the slow solar wind. These observational trends indicate that the occurrence of mass-proportional temperature ratios is closely associated with weakly collisional plasma conditions, whereas near-equal proton and alpha-particle temperatures are preferentially observed in more collisionally processed solar wind.

The observed relationship between $T_\alpha/T_p$ and collisional age is broadly consistent with the established role of Coulomb collisions in regulating non-equilibrium ion properties during solar wind expansion. Since the collisional age scales approximately as $A_c \propto n_p/(v_{sw}T_p^{3/2})$ \citep{Kasper2008}, slow solar wind, which is generally denser and cooler, accumulates more Coulomb collisions than fast solar wind. Consequently, collisions progressively reduce temperature differences between ion species, leading to distributions dominated by $T_\alpha/T_p \sim 1$ in the slow solar wind. In contrast, the low collisional age of the fast solar wind allows elevated alpha-to-proton temperature ratios to persist over heliocentric distances, resulting in the predominance of the $T_\alpha/T_p \sim 4$ population at 1 AU. This interpretation is consistent with previous observational studies showing that alpha-particle differential heating and differential streaming decrease with increasing collisional processing \citep{Kasper2008, Maruca2013, Alterman2018, Tracy2015, Durovcova2021}.

However, the present observations alone cannot distinguish whether the elevated $T_\alpha/T_p$ values observed at 1 AU result from continued local heating during solar wind expansion or simply represent the preservation of heating signatures established closer to the Sun. Instead, our results demonstrate that high temperature ratios are preferentially associated with weakly collisional plasma, whereas near-equal proton and alpha-particle temperatures occur predominantly in more collisionally processed plasma. This distinction is important because collisional age primarily governs the degree to which non-equilibrium signatures are preserved rather than directly identifying the physical mechanism responsible for generating them.

The preferential heating itself is widely believed to originate much closer to the Sun, where the plasma first becomes effectively collisionless \citep{Kohl1998, Cranmer2009, Landi2009, Ofman2010, Chandran2013, Mostafavi2024, Martinovic2025, Mihailo2026}. Remote spectroscopic observations indicate that heavy ions become substantially hotter than protons only a few tenths of a solar radius above the photosphere as the Coulomb collision frequency decreases rapidly with height \citep{Kohl1998, Landi2009, Cranmer2009}. Parker Solar Probe measurements have likewise revealed strong proton and alpha-particle temperature anisotropies, enhanced Alfv\'enic fluctuations, and weak collisional coupling in the near-Sun solar wind \citep{Mostafavi2024, Yogesh2026}. These observations are broadly consistent with theoretical models in which wave--particle interactions, turbulent dissipation, stochastic heating, ion-cyclotron resonance, and drift instabilities provide the free energy required for preferential ion heating \citep{Chandran2010, Chandran2013, Cranmer2012, Martinovic2020, Martinovic2025, Bowen2025, Mihailo2026}. While our measurements at 1 AU cannot directly determine which of these processes dominates, they are consistent with a scenario in which preferential heating is established in the near-Sun corona and inner heliosphere and subsequently modified by Coulomb collisions during outward expansion.

Our results therefore support a picture in which the observed alpha-to-proton temperature ratio at 1 AU reflects the combined effects of kinetic heating near the Sun and collisional relaxation during heliospheric transport. The relative importance of these two processes depends strongly on the collisional history of the plasma, naturally explaining why mass-proportional temperature ratios are most commonly observed in weakly collisional solar wind, whereas nearly equal ion temperatures dominate in more collisionally processed intervals.

In addition to the strong dependence on collisional age, our results reveal a clear solar-cycle modulation in the occurrence of both weakly collisional plasma ($A_c < 1$) and mass-proportional alpha-particle temperatures ($T_\alpha/T_p > 4$). In the slow solar wind ($<300$ and $300$--$400$ km s$^{-1}$), both quantities increase during solar maxima, whereas the intermediate ($400$--$500$ km s$^{-1}$) and fast ($>500$ km s$^{-1}$) solar wind exhibit the opposite behavior, with larger occurrence fractions during solar minima. These opposite correlations indicate that the occurrence of mass-proportional temperature ratios depends not only on solar wind speed but also on the evolving distribution of solar wind plasma sampled throughout the solar cycle. It is worth noting that the interpretation of the slowest solar wind ($<300$ km s$^{-1}$) requires some caution. Previous studies have shown that this speed range may be affected by instrumental limitations or sampling biases, and is therefore often excluded or treated separately in statistical analyses \citep{Kasper2007, Alterman2019, Alterman2025}. Nevertheless, the systematic trends identified here persist across the broader slow-wind regime ($300$--$400$ km s$^{-1}$), indicating that the observed solar-cycle dependence is not driven solely by the lowest-speed intervals.

The interpretation of these trends, however, requires caution because neither collisional age nor solar wind speed uniquely identifies the solar wind source region. Previous studies have shown that collisional age ($A_c$) and proton specific entropy ($S_p \propto \log[T_p^{3/2}/n_p]$) are closely related through their common dependence on the Coulomb collision frequency ($\nu_{p} \propto n_pT_p^{-3/2}$), such that $-\log(\nu_{pp})$ scales with $S_p$ up to an additive constant \citep{Ben2019, Heidrich2020, Alterman2018, Xu2015}. Consequently, plasma with low collisional age ($A_c < 1$) generally also exhibits high specific entropy, indicating that it has experienced relatively little collisional processing during its expansion from the Sun. Proton specific entropy has been identified as one of several useful classifiers of solar wind source regions rather than a unique discriminator \citep{Xu2015}. Likewise, solar wind speed alone does not uniquely distinguish plasma originating from coronal holes, active regions, or streamer-belt sources, particularly in the intermediate-speed regime where multiple source populations coexist \citep{DAmicis2021, Alterman2025b}. Consequently, the present analysis does not attempt to assign a unique source region to individual intervals but instead examines the statistical evolution of plasma populations observed near the ecliptic over nearly three solar cycles.


A plausible interpretation is that the observed solar-cycle dependence reflects changes in the relative contributions of different solar wind source populations reaching 1 AU. During solar minima, the near-ecliptic heliosphere is more frequently occupied by long-lived high-speed streams originating from open magnetic-field regions associated with coronal holes \citep{McComas2003, McComas2008, Wang2009}. These streams typically exhibit lower densities, higher proton temperatures, stronger Alfv\'enicity, lower collisional ages, and enhanced non-equilibrium ion signatures, allowing preferential heating established closer to the Sun to remain visible at 1 AU \citep{DAmicis2021, Rivera2024, Alterman2025b}. In contrast, during solar maxima the ecliptic is increasingly populated by slower and more variable solar wind associated with streamer belts, active regions, and the higher occurrence of transient coronal mass ejections (CMEs) \citep{Antiochos2011, Abbo2016, Yardley2024, Rivera2025a, Rivera2025b}. Such plasma generally experiences greater collisional processing during its expansion and consequently exhibits a larger fraction of intervals with nearly equal proton and alpha-particle temperatures. The strongest solar-cycle variability is observed in the solar wind of speed from 300-500 km s$^{-1}$, where plasma originating from coronal holes and closed-field regions is expected to coexist, suggesting that the observed trends primarily reflect changes in the relative occurrence of these source populations sampled by Wind near the ecliptic over the solar cycle. By contrast, the comparatively weak correlations in the slowest ($<300$ km s$^{-1}$) and fastest ($>500$ km s$^{-1}$) solar wind suggest that the observed long-term variations should be interpreted cautiously. In particular, the weak correlations in the fastest ($>500$ km s$^{-1}$) solar wind are consistent with this regime being dominated by relatively homogeneous coronal-hole wind, whose properties vary less over the solar cycle.

The intermediate-speed ($400-500$ km s$^{-1}$) solar wind deserves particular attention because it exhibits the strongest reversal in the solar-cycle dependence. Rather than representing a distinct plasma state, this speed range is widely recognized as a mixture of solar wind originating from multiple coronal environments, including coronal-hole boundaries, active-region outflows, and streamer-associated plasma \citep{DAmicis2021, Rivera2024, Alterman2025b, Alterman2026}. The observed reversal near $400-500$ km s$^{-1}$ therefore likely reflects changes in the relative abundance of these source populations over the solar cycle, rather than a fundamental transition occurring at a single solar wind speed. This interpretation is consistent with the gradual evolution of the fitted population fractions presented in Figures~2 and 4, which show no evidence for an abrupt boundary between slow and fast solar wind.

Overall, our results suggest that the alpha-to-proton temperature ratio observed at 1 AU is controlled by the combined influence of preferential heating established in the inner heliosphere, subsequent collisional relaxation during solar wind expansion, and the changing mixture of solar wind source populations sampled throughout the solar cycle. Future studies combining in-situ plasma measurements with compositional diagnostics, magnetic connectivity, and remote observations of coronal source regions will be required to distinguish more directly between these contributing effects.

The present study provides the long-term statistical investigations of the combined dependence of the alpha-to-proton temperature ratio on solar wind speed, collisional age, and solar-cycle phase using nearly three solar cycles of Wind observations. While previous studies have established that preferential ion heating is more prominent in weakly collisional solar wind \citep{Kasper2008, Maruca2013, Alterman2018}, our analysis demonstrates that the relative occurrence of the equal-temperature and mass-proportional plasma populations also evolves systematically with solar activity. These results indicate that solar-cycle variability should be considered when interpreting statistical studies of ion heating and collisional evolution at 1 AU.

Future investigations combining in-situ plasma measurements from Parker Solar Probe, Solar Orbiter, Wind, ACE, and Aditya-L1 with remote observations of coronal magnetic topology and solar wind source regions will provide a more direct means of connecting the evolution of $T_\alpha/T_p$ to the physical processes operating close to the Sun. In particular, coordinated multi-spacecraft observations spanning a wide range of heliocentric distances will enable the radial evolution of preferential ion heating and collisional relaxation to be tracked continuously from the corona to 1 AU. Such studies will help determine how kinetic heating, collisional processing, and the changing mixture of solar wind source populations collectively shape the thermodynamic evolution of solar wind ions throughout the heliosphere.

\section{Conclusion}
Using nearly three solar cycles of Wind observations, we investigated the dependence of the alpha-to-proton temperature ratio ($T_\alpha/T_p$) on solar wind speed, collisional age, and solar-cycle phase. The observed distributions reveal two dominant populations corresponding to near-equal temperatures ($T_\alpha/T_p \sim 1$) and mass-proportional temperatures ($T_\alpha/T_p \sim 4$), whose relative occurrence varies systematically with solar wind speed. The slow solar wind is dominated by more collisionally processed plasma with larger collisional ages and temperature ratios close to unity, whereas the fast solar wind exhibits systematically lower collisional ages and persistent preferential alpha-particle heating with $T_\alpha/T_p \gtrsim 4$. The occurrence of mass-proportional temperature ratios also shows a clear solar-cycle dependence that varies across different solar wind speed regimes. In the slow wind, mass-proportional temperature ratios become more common during solar maxima, whereas in the intermediate and high-speed wind they are enhanced during solar minima.

These results indicate that the observed alpha-to-proton temperature ratio at 1 AU reflects the combined influence of preferential ion heating established closer to the Sun, subsequent Coulomb collisional relaxation during solar wind expansion, and the evolving mixture of solar wind source populations sampled over the solar cycle. While the present analysis cannot uniquely distinguish between ongoing local heating and the preservation of near-Sun heating signatures, the observed trends are consistent with weaker collisional processing and stronger preferential heating in fast solar wind, and enhanced thermalization in slower, denser solar wind. Overall, this work provides new observational constraints on the interplay between preferential ion heating, collisional evolution, and solar-cycle variability, thereby improving our understanding of the thermodynamic evolution of heavy ions in the solar wind plasma.

\begin{acknowledgments}
We gratefully acknowledge the principal investigators of Solar Wind Experiment (SWE) on-board NASA's Wind mission for generating the data and making them publicly available. L.O. and Y. acknowledge support from NASA grant 80NSSC24K0724, NSF grant AGS-2300961 and NASA’s GSFC through Cooperative Agreement 80NSSC21M0180 to the Catholic University of America, Partnership for Heliophysics and Space Environment Research (PHaSER). Y. acknowledge the support by the College of Liberal Arts and Sciences at the University of Iowa. G.G.H. acknowledges support of NASA grant 80NSSC24K124. We also thank World Data Center–SILSO, Royal Observatory of Belgium, Brussels for the monthly mean sunspot number data used in this paper. We express our sincere gratitude to the Department of Space, Government of India for supporting this work. 

\textit{Facility:} Wind Spacecraft (Solar Wind Experiment)
\end{acknowledgments}

\section*{DATA AVAILABILITY STATEMENTS} 

All data used in this study are publicly available. The solar wind bulk plasma parameters were obtained from NASA’s CDAWeb portal (\url{https://cdaweb.gsfc.nasa.gov/}
) under DOI: \doi{10.48322/nasd-j276}. The Wind bow shock database is accessible at \url{https://wind.nasa.gov/mfi/bow_shock.html}
. The monthly mean sunspot number data were provided by WDC–SILSO, Royal Observatory of Belgium, Brussels (\url{https://www.sidc.be/SILSO/datafiles}
), under DOI: \doi{10.24414/qnza-ac80}.

\bibliography{main}{}

@ARTICLE{Parker1958,
       author = {{Parker}, E.~N.},
        title = "{Dynamics of the Interplanetary Gas and Magnetic Fields.}",
      journal = {\apj},
         year = 1958,
        month = nov,
       volume = {128},
        pages = {664},
          doi = {10.1086/146579},
       adsurl = {https://ui.adsabs.harvard.edu/abs/1958ApJ...128..664P}
}

@article{vonSteiger2000,
       author = {{von Steiger}, R. and {Schwadron}, N.~A. and {Fisk}, L.~A. and {Geiss}, J. and {Gloeckler}, G. and {Hefti}, S. and {Wilken}, B. and {Wimmer-Schweingruber}, R.~F. and {Zurbuchen}, T.~H.},
        title = "{Composition of quasi-stationary solar wind flows from Ulysses/Solar Wind Ion Composition Spectrometer}",
      journal = {\jgr},
         year = 2000,
        month = dec,
       volume = {105},
       number = {A12},
        pages = {27217-27238},
          doi = {10.1029/1999JA000358},
       adsurl = {https://ui.adsabs.harvard.edu/abs/2000JGR...10527217V}
}

@article{Marsch2006,
       author = {{Marsch}, Eckart},
        title = "{Kinetic Physics of the Solar Corona and Solar Wind}",
      journal = {Living Reviews in Solar Physics},
         year = 2006,
        month = dec,
       volume = {3},
       number = {1},
          eid = {1},
        pages = {1},
          doi = {10.12942/lrsp-2006-1},
       adsurl = {https://ui.adsabs.harvard.edu/abs/2006LRSP....3....1M}
}

@ARTICLE{Yogesh2021,
       author = {{Yogesh} and {Chakrabarty}, D. and {Srivastava}, N.},
        title = "{Evidence for distinctive changes in the solar wind helium abundance in solar cycle 24}",
      journal = {\mnras},
         year = 2021,
        month = may,
       volume = {503},
       number = {1},
        pages = {L17-L22},
          doi = {10.1093/mnrasl/slab016},
archivePrefix = {arXiv},
       eprint = {2102.05395},
 primaryClass = {astro-ph.SR},
       adsurl = {https://ui.adsabs.harvard.edu/abs/2021MNRAS.503L..17Y}
}

@ARTICLE{Yogesh2022,
       author = {Yogesh and Chakrabarty, D. and Srivastava, N.},
        title = "{A holistic approach to understand helium enrichment in interplanetary coronal mass ejections: new insights}",
      journal = {Monthly Notices of the Royal Astronomical Society: Letters},
         year = 2022,
        month = jun,
       volume = {513},
       number = {1},
        pages = {L106-L111},
          doi = {10.1093/mnrasl/slac044},
archivePrefix = {arXiv},
       eprint = {2202.01722},
 primaryClass = {astro-ph.SR},
       adsurl = {https://ui.adsabs.harvard.edu/abs/2022MNRAS.513L.106Y}
}

@ARTICLE{Yogesh2023,
       author = {Yogesh and Chakrabarty, D. and Srivastava, Nandita},
        title = "{New insights on the behaviour of solar wind protons and alphas in the stream interaction region in solar cycle 23 and 24}",
      journal = {Monthly Notices of the Royal Astronomical Society: Letters},
         year = 2023,
        month = nov,
       volume = {526},
       number = {1},
        pages = {L13-L19},
          doi = {10.1093/mnrasl/slad112},
archivePrefix = {arXiv},
       eprint = {2304.00274},
 primaryClass = {astro-ph.SR},
       adsurl = {https://ui.adsabs.harvard.edu/abs/2023MNRAS.526L..13Y}
}

@article{Ofman2004,
author = {Ofman, L.},
title = {Three-fluid model of the heating and acceleration of the fast solar wind},
journal = {Journal of Geophysical Research: Space Physics},
volume = {109},
number = {A7},
pages = {},
doi = {https://doi.org/10.1029/2003JA010221},
url = {https://agupubs.onlinelibrary.wiley.com/doi/abs/10.1029/2003JA010221},
eprint = {https://agupubs.onlinelibrary.wiley.com/doi/pdf/10.1029/2003JA010221},
year = {2004}
}

@article{Ofman2022,
  title={Modeling ion beams, kinetic instabilities, and waves observed by the Parker Solar Probe near perihelia},
  author={Ofman, Leon and Boardsen, Scott A and Jian, Lan K and Verniero, Jaye L and Larson, Davin},
  journal={The Astrophysical Journal},
  volume={926},
  number={2},
  pages={185},
  year={2022},
  doi = {10.3847/1538-4357/ac402c},
  publisher={The American Astronomical Society}
}

@article{Ofman2023,
  title={Observations and modeling of unstable proton and $\alpha$ particle velocity distributions in Sub-Alfv{\'e}nic solar wind at parker solar probe perihelia},
  author={Ofman, Leon and Boardsen, Scott A and Jian, Lan K and Mostafavi, Parisa and Verniero, Jaye L and Livi, Roberto and McManus, Michael and Rahmati, Ali and Larson, Davin and Stevens, Michael L},
  journal={The Astrophysical Journal},
  volume={954},
  number={2},
  pages={109},
  year={2023},
  doi = {10.3847/1538-4357/acea7e},
  publisher={The American Astronomical Society}
}

@ARTICLE{Ofman2024,
       author = {{Ofman}, Leon and {Yogesh} and {Giordano}, Silvio},
        title = "{Understanding the Variability of Helium Abundance in the Solar Corona Using Three-fluid Modeling and Ultraviolet Observations}",
      journal = {\apjl},
         year = 2024,
        month = jul,
       volume = {970},
       number = {1},
          eid = {L16},
        pages = {L16},
          doi = {10.3847/2041-8213/ad5e7e},
archivePrefix = {arXiv},
       eprint = {2406.14897},
 primaryClass = {astro-ph.SR},
       adsurl = {https://ui.adsabs.harvard.edu/abs/2024ApJ...970L..16O}
}

@ARTICLE{Yogesh2024,
       author = {{Yogesh} and {Gopalswamy}, N. and {Chakrabarty}, D. and {Mostafavi}, Parisa and {Yashiro}, Seiji and {Srivastava}, Nandita and {Ofman}, Leon},
        title = "{Origins of Very Low Helium Abundance Streams Detected in the Solar Wind Plasma}",
      journal = {\apj},
         year = 2024,
        month = dec,
       volume = {977},
       number = {1},
          eid = {89},
        pages = {89},
          doi = {10.3847/1538-4357/ad84d6},
archivePrefix = {arXiv},
       eprint = {2410.04713},
 primaryClass = {astro-ph.SR},
       adsurl = {https://ui.adsabs.harvard.edu/abs/2024ApJ...977...89Y}
}

@book{Bame1972,
       author = {{Bame}, S.~J.},
        title = "{Spacecraft Observations of the Solar Wind Composition}",
    booktitle = {NASA Special Publication},
         year = 1972,
       editor = {{Sonett}, Charles P. and {Coleman}, Paul Jerome and {Wilcox}, John Marsh},
       volume = {308},
        pages = {535},
       adsurl = {https://ui.adsabs.harvard.edu/abs/1972NASSP.308..535B}
}

@ARTICLE{Marsch2012,
       author = {{Marsch}, Eckart},
        title = "{Helios: Evolution of Distribution Functions 0.3-1 AU}",
      journal = {\ssr},
         year = 2012,
        month = nov,
       volume = {172},
       number = {1-4},
        pages = {23-39},
          doi = {10.1007/s11214-010-9734-z},
       adsurl = {https://ui.adsabs.harvard.edu/abs/2012SSRv..172...23M}
}

@article{Verscharen2019,
       author = {{Verscharen}, Daniel and {Klein}, Kristopher G. and {Maruca}, Bennett A.},
        title = "{The multi-scale nature of the solar wind}",
      journal = {Living Reviews in Solar Physics},
         year = 2019,
        month = dec,
       volume = {16},
       number = {1},
          eid = {5},
        pages = {5},
          doi = {10.1007/s41116-019-0021-0},
archivePrefix = {arXiv},
       eprint = {1902.03448},
 primaryClass = {physics.space-ph},
       adsurl = {https://ui.adsabs.harvard.edu/abs/2019LRSP...16....5V}
}

@ARTICLE{Ryan1975,
       author = {{Ryan}, J.~M. and {Axford}, W.~I.},
        title = "{The behaviour of minor species in the solar wind.}",
      journal = {Journal of Geophysics Zeitschrift Geophysik},
         year = 1975,
        month = jan,
       volume = {41},
        pages = {221-232},
       adsurl = {https://ui.adsabs.harvard.edu/abs/1975JGZG...41..221R}
}

@ARTICLE{Marsch1982,
       author = {{Marsch}, E. and {Goertz}, C.~K. and {Richter}, K.},
        title = "{Wave heating and acceleration of solar wind ions by cyclotron resonance}",
      journal = {\jgr},
         year = 1982,
        month = jul,
       volume = {87},
       number = {A7},
        pages = {5030-5044},
          doi = {10.1029/JA087iA07p05030},
       adsurl = {https://ui.adsabs.harvard.edu/abs/1982JGR....87.5030M}
}

@techreport{vonSteiger1995,
  title={Solar wind ion composition and charge states},
  author={Vonsteiger, R},
  year={1995},
  institution={California Institute of Technology (CalTech), Pasadena, CA (United States~…}
}

@article{Durovcova2019,
       author = {{Durovcov{\'a}}, Tereza and {{\v{S}}afr{\'a}nkov{\'a}}, Jana and {N{\v{e}}me{\v{c}}ek}, Zden{\v{e}}k},
        title = "{Evolution of Relative Drifts in the Expanding Solar Wind: Helios Observations}",
      journal = {\solphys},
         year = 2019,
        month = jul,
       volume = {294},
       number = {7},
          eid = {97},
        pages = {97},
          doi = {10.1007/s11207-019-1490-y},
       adsurl = {https://ui.adsabs.harvard.edu/abs/2019SoPh..294...97D}
}

@article{Verniero2022,
  title={Strong perpendicular velocity-space diffusion in proton beams observed by parker solar probe},
  author={Verniero, JL and Chandran, BDG and Larson, DE and Paulson, K and Alterman, BL and Badman, S and Bale, SD and Bonnell, JW and Bowen, TA and de Wit, T Dudok and others},
  journal={The Astrophysical Journal},
  volume={924},
  number={2},
  pages={112},
  year={2022},
  doi = {10.3847/1538-4357/ac36d5},
  publisher={The American Astronomical Society}
}

@ARTICLE{Mostafavi2022,
       author = {{Mostafavi}, P. and {Allen}, R.~C. and {McManus}, M.~D. and {Ho}, G.~C. and {Raouafi}, N.~E. and {Larson}, D.~E. and {Kasper}, J.~C. and {Bale}, S.~D.},
        title = "{Alpha-Proton Differential Flow of the Young Solar Wind: Parker Solar Probe Observations}",
      journal = {\apjl},
         year = 2022,
        month = feb,
       volume = {926},
       number = {2},
          eid = {L38},
        pages = {L38},
          doi = {10.3847/2041-8213/ac51e1},
archivePrefix = {arXiv},
       eprint = {2202.03551},
 primaryClass = {physics.space-ph},
       adsurl = {https://ui.adsabs.harvard.edu/abs/2022ApJ...926L..38M}
}

@article{Cranmer2009,
       author = {{Cranmer}, Steven R. and {Matthaeus}, William H. and {Breech}, Benjamin A. and {Kasper}, Justin C.},
        title = "{Empirical Constraints on Proton and Electron Heating in the Fast Solar Wind}",
      journal = {\apj},
         year = 2009,
        month = sep,
       volume = {702},
       number = {2},
        pages = {1604-1614},
          doi = {10.1088/0004-637X/702/2/1604},
archivePrefix = {arXiv},
       eprint = {0907.2650},
 primaryClass = {astro-ph.SR},
       adsurl = {https://ui.adsabs.harvard.edu/abs/2009ApJ...702.1604C}
}

@article{Landi2009,
       author = {{Landi}, E. and {Cranmer}, S.~R.},
        title = "{Ion Temperatures in the Low Solar Corona: Polar Coronal Holes at Solar Minimum}",
      journal = {\apj},
         year = 2009,
        month = jan,
       volume = {691},
       number = {1},
        pages = {794-805},
          doi = {10.1088/0004-637X/691/1/794},
archivePrefix = {arXiv},
       eprint = {0810.0018},
 primaryClass = {astro-ph},
       adsurl = {https://ui.adsabs.harvard.edu/abs/2009ApJ...691..794L}
}

@article{Tracy2016,
       author = {{Tracy}, Patrick J. and {Kasper}, Justin C. and {Raines}, Jim M. and {Shearer}, Paul and {Gilbert}, Jason A. and {Zurbuchen}, Thomas H.},
        title = "{Constraining Solar Wind Heating Processes by Kinetic Properties of Heavy Ions}",
      journal = {\prl},
         year = 2016,
        month = jun,
       volume = {116},
       number = {25},
          eid = {255101},
        pages = {255101},
          doi = {10.1103/PhysRevLett.116.255101},
       adsurl = {https://ui.adsabs.harvard.edu/abs/2016PhRvL.116y5101T}
}

@article{Isenberg2009,
       author = {{Isenberg}, Philip A. and {Vasquez}, Bernard J.},
        title = "{Preferential Acceleration and Perpendicular Heating of Minor Ions in a Collisionless Coronal Hole}",
      journal = {\apj},
         year = 2009,
        month = may,
       volume = {696},
       number = {1},
        pages = {591-600},
          doi = {10.1088/0004-637X/696/1/591},
       adsurl = {https://ui.adsabs.harvard.edu/abs/2009ApJ...696..591I}
}

@article{Jian2010,
       author = {{Jian}, L.~K. and {Russell}, C.~T. and {Luhmann}, J.~G. and {Anderson}, B.~J. and {Boardsen}, S.~A. and {Strangeway}, R.~J. and {Cowee}, M.~M. and {Wennmacher}, A.},
        title = "{Observations of ion cyclotron waves in the solar wind near 0.3 AU}",
      journal = {Journal of Geophysical Research (Space Physics)},
         year = 2010,
        month = dec,
       volume = {115},
       number = {A12},
          eid = {A12115},
        pages = {A12115},
          doi = {10.1029/2010JA015737},
       adsurl = {https://ui.adsabs.harvard.edu/abs/2010JGRA..11512115J}
}

@ARTICLE{McIntosh2011,
       author = {{McIntosh}, Scott W. and {de Pontieu}, Bart and {Carlsson}, Mats and {Hansteen}, Viggo and {Boerner}, Paul and {Goossens}, Marcel},
        title = "{Alfv{\'e}nic waves with sufficient energy to power the quiet solar corona and fast solar wind}",
      journal = {\nat},
         year = 2011,
        month = jul,
       volume = {475},
       number = {7357},
        pages = {477-480},
          doi = {10.1038/nature10235},
       adsurl = {https://ui.adsabs.harvard.edu/abs/2011Natur.475..477M}
}

@article{Kasper2013,
       author = {{Kasper}, Justin C. and {Maruca}, Bennett A. and {Stevens}, Michael L. and {Zaslavsky}, Arnaud},
        title = "{Sensitive Test for Ion-Cyclotron Resonant Heating in the Solar Wind}",
      journal = {\prl},
         year = 2013,
        month = mar,
       volume = {110},
       number = {9},
          eid = {091102},
        pages = {091102},
          doi = {10.1103/PhysRevLett.110.091102},
       adsurl = {https://ui.adsabs.harvard.edu/abs/2013PhRvL.110i1102K}
}

@article{Navarro2020,
       author = {{Navarro}, Roberto E. and {Mu{\~n}oz}, V{\'\i}ctor and {Valdivia}, Juan A. and {Moya}, Pablo S.},
        title = "{Feasibility of Ion-cyclotron Resonant Heating in the Solar Wind}",
      journal = {\apjl},
         year = 2020,
        month = jul,
       volume = {898},
       number = {1},
          eid = {L9},
        pages = {L9},
          doi = {10.3847/2041-8213/aba0ae},
       adsurl = {https://ui.adsabs.harvard.edu/abs/2020ApJ...898L...9N}
}

@ARTICLE{Ofman2025,
       author = {{Ofman}, L. and {Yogesh} and {Boardsen}, S.~A. and {Mostafavi}, P. and {Jian}, L.~K. and {Sadykov}, V.~M. and {Klein}, K. and {Martinovic}, M.},
        title = "{Modeling Hot, Anisotropic Ion Beams in the Solar Wind Motivated by the Parker Solar Probe Observations near Perihelia.}",
      journal = {\apj},
         year = 2025,
        month = jan,
       volume = {984},
          eid = {174},
        pages = {174},
          doi = {https://doi.org/10.3847/1538-4357/adc812},
archivePrefix = {arXiv},
       eprint = {2504.00659},
 primaryClass = {astro-ph.SR},
       adsurl = {https://ui.adsabs.harvard.edu/abs/2025ApJ...984..174O}
}

@ARTICLE{Yogesh2025,
       author = {{Yogesh} and {Ofman}, Leon and {Boardsen}, Scott A. and {Klein}, Kristopher and {Martinovi{\'c}}, Mihailo and {Sadykov}, Viacheslav M. and {Verniero}, Jaye L. and {Shankarappa}, Niranjana and {Jian}, Lan K. and {Mostafavi}, Parisa and {Huang}, Jia and {Paulson}, K.~W.},
        title = "{Evidence of Interaction between Ion-scale Waves and Ion Velocity Distributions in the Solar Wind}",
      journal = {\apj},
         year = 2025,
        month = jun,
       volume = {986},
       number = {2},
          eid = {119},
        pages = {119},
          doi = {10.3847/1538-4357/add467},
archivePrefix = {arXiv},
       eprint = {2505.02999},
 primaryClass = {astro-ph.SR},
       adsurl = {https://ui.adsabs.harvard.edu/abs/2025ApJ...986..119Y}
}

@article{Chandran2010,
       author = {{Chandran}, Benjamin D.~G. and {Li}, Bo and {Rogers}, Barrett N. and {Quataert}, Eliot and {Germaschewski}, Kai},
        title = "{Perpendicular Ion Heating by Low-frequency Alfv{\'e}n-wave Turbulence in the Solar Wind}",
      journal = {\apj},
         year = 2010,
        month = sep,
       volume = {720},
       number = {1},
        pages = {503-515},
          doi = {10.1088/0004-637X/720/1/503},
archivePrefix = {arXiv},
       eprint = {1001.2069},
 primaryClass = {astro-ph.SR},
       adsurl = {https://ui.adsabs.harvard.edu/abs/2010ApJ...720..503C}
}

@article{Chandran2013,
       author = {{Chandran}, B.~D.~G. and {Verscharen}, D. and {Quataert}, E. and {Kasper}, J.~C. and {Isenberg}, P.~A. and {Bourouaine}, S.},
        title = "{Stochastic Heating, Differential Flow, and the Alpha-to-proton Temperature Ratio in the Solar Wind}",
      journal = {\apj},
         year = 2013,
        month = oct,
       volume = {776},
       number = {1},
          eid = {45},
        pages = {45},
          doi = {10.1088/0004-637X/776/1/45},
archivePrefix = {arXiv},
       eprint = {1307.8090},
 primaryClass = {astro-ph.SR},
       adsurl = {https://ui.adsabs.harvard.edu/abs/2013ApJ...776...45C}
}

@article{Verscharen2013,
       author = {{Verscharen}, Daniel and {Bourouaine}, Sofiane and {Chandran}, Benjamin D.~G.},
        title = "{Instabilities Driven by the Drift and Temperature Anisotropy of Alpha Particles in the Solar Wind}",
      journal = {\apj},
         year = 2013,
        month = aug,
       volume = {773},
       number = {2},
          eid = {163},
        pages = {163},
          doi = {10.1088/0004-637X/773/2/163},
archivePrefix = {arXiv},
       eprint = {1307.1823},
 primaryClass = {physics.space-ph},
       adsurl = {https://ui.adsabs.harvard.edu/abs/2013ApJ...773..163V}
}

@article{Cargill2004,
       author = {{Cargill}, Peter J. and {Klimchuk}, James A.},
        title = "{Nanoflare Heating of the Corona Revisited}",
      journal = {\apj},
         year = 2004,
        month = apr,
       volume = {605},
       number = {2},
        pages = {911-920},
          doi = {10.1086/382526},
       adsurl = {https://ui.adsabs.harvard.edu/abs/2004ApJ...605..911C}
}

@article{Drake2009,
       author = {{Drake}, J.~F. and {Cassak}, P.~A. and {Shay}, M.~A. and {Swisdak}, M. and {Quataert}, E.},
        title = "{A Magnetic Reconnection Mechanism for Ion Acceleration and Abundance Enhancements in Impulsive Flares}",
      journal = {\apjl},
         year = 2009,
        month = jul,
       volume = {700},
       number = {1},
        pages = {L16-L20},
          doi = {10.1088/0004-637X/700/1/L16},
       adsurl = {https://ui.adsabs.harvard.edu/abs/2009ApJ...700L..16D}
}

@article{Artemyev2014,
       author = {{Artemyev}, A.~V. and {Zimbardo}, G. and {Ukhorskiy}, A.~Y. and {Fujimoto}, M.},
        title = "{Preferential acceleration of heavy ions in the reconnection outflow region. Drift and surfatron ion acceleration}",
      journal = {\aap},
         year = 2014,
        month = feb,
       volume = {562},
          eid = {A58},
        pages = {A58},
          doi = {10.1051/0004-6361/201322462},
       adsurl = {https://ui.adsabs.harvard.edu/abs/2014A&A...562A..58A}
}

@article{Duan2023,
       author = {{Duan}, Yadan and {Shen}, Yuandeng and {Chen}, Hechao and {Tang}, Zehao and {Zhou}, Chenrui and {Zhou}, Xinping and {Tan}, Song},
        title = "{Macrospicules and Their Connection to Magnetic Reconnection in the Lower Solar Atmosphere}",
      journal = {\apjl},
         year = 2023,
        month = jan,
       volume = {942},
       number = {1},
          eid = {L22},
        pages = {L22},
          doi = {10.3847/2041-8213/acac2b},
archivePrefix = {arXiv},
       eprint = {2212.03425},
 primaryClass = {astro-ph.SR},
       adsurl = {https://ui.adsabs.harvard.edu/abs/2023ApJ...942L..22D}
}

@INPROCEEDINGS{Scudder1992,
       author = {{Scudder}, J.~D.},
        title = "{The cause of the coronal temperature inversion of the solar atmosphere and the implications for the solar wind}",
    booktitle = {Solar Wind Seven Colloquium},
         year = 1992,
       editor = {{Marsch}, E. and {Schwenn}, R.},
        month = jan,
        pages = {103-112},
       adsurl = {https://ui.adsabs.harvard.edu/abs/1992sws..coll..103S}
}

@article{Chandran2010b,
       author = {{Chandran}, Benjamin D.~G.},
        title = "{Alfv{\'e}n-wave Turbulence and Perpendicular Ion Temperatures in Coronal Holes}",
      journal = {\apj},
         year = 2010,
        month = sep,
       volume = {720},
       number = {1},
        pages = {548-554},
          doi = {10.1088/0004-637X/720/1/548},
archivePrefix = {arXiv},
       eprint = {1006.3473},
 primaryClass = {astro-ph.SR},
       adsurl = {https://ui.adsabs.harvard.edu/abs/2010ApJ...720..548C}
}

@article{Kasper2008,
       author = {{Kasper}, J.~C. and {Lazarus}, A.~J. and {Gary}, S.~P.},
        title = "{Hot Solar-Wind Helium: Direct Evidence for Local Heating by Alfv{\'e}n-Cyclotron Dissipation}",
      journal = {\prl},
         year = 2008,
        month = dec,
       volume = {101},
       number = {26},
          eid = {261103},
        pages = {261103},
          doi = {10.1103/PhysRevLett.101.261103},
       adsurl = {https://ui.adsabs.harvard.edu/abs/2008PhRvL.101z1103K}
}

@article{Kasper2017,
       author = {{Kasper}, J.~C. and {Klein}, K.~G. and {Weber}, T. and {Maksimovic}, M. and {Zaslavsky}, A. and {Bale}, S.~D. and {Maruca}, B.~A. and {Stevens}, M.~L. and {Case}, A.~W.},
        title = "{A Zone of Preferential Ion Heating Extends Tens of Solar Radii from the Sun}",
      journal = {\apj},
         year = 2017,
        month = nov,
       volume = {849},
       number = {2},
          eid = {126},
        pages = {126},
          doi = {10.3847/1538-4357/aa84b1},
archivePrefix = {arXiv},
       eprint = {1708.05683},
 primaryClass = {astro-ph.SR},
       adsurl = {https://ui.adsabs.harvard.edu/abs/2017ApJ...849..126K}
}

@article{Maruca2013,
       author = {{Maruca}, B.~A. and {Bale}, S.~D. and {Sorriso-Valvo}, L. and {Kasper}, J.~C. and {Stevens}, M.~L.},
        title = "{Collisional Thermalization of Hydrogen and Helium in Solar-Wind Plasma}",
      journal = {\prl},
         year = 2013,
        month = dec,
       volume = {111},
       number = {24},
          eid = {241101},
        pages = {241101},
          doi = {10.1103/PhysRevLett.111.241101},
archivePrefix = {arXiv},
       eprint = {1311.5473},
 primaryClass = {physics.space-ph},
       adsurl = {https://ui.adsabs.harvard.edu/abs/2013PhRvL.111x1101M}
}

@article{Feldman1974,
       author = {{Feldman}, W.~C. and {Asbridge}, J.~R. and {Bame}, S.~J.},
        title = "{The solar wind He$^{2+}$ to H$^{+}$ temperature ratio}",
      journal = {\jgr},
         year = 1974,
        month = jan,
       volume = {79},
       number = {16},
        pages = {2319},
          doi = {10.1029/JA079i016p02319},
       adsurl = {https://ui.adsabs.harvard.edu/abs/1974JGR....79.2319F}
}

@article{Neugebauer1976,
    author = {{Neugebauer}, M.},
        title = "{The role of Coulomb collisions in limiting differential flow and temperature differences in the solar wind}",
      journal = {\jgr},
         year = 1976,
        month = jan,
       volume = {81},
       number = {1},
        pages = {78},
          doi = {10.1029/JA081i001p00078},
       adsurl = {https://ui.adsabs.harvard.edu/abs/1976JGR....81...78N}
}

@article{Livi1986,
       author = {{Livi}, S. and {Marsch}, E. and {Rosenbauer}, H.},
        title = "{Coulomb collisional domains in solar wind}",
      journal = {\jgr},
         year = 1986,
        month = jul,
       volume = {91},
       number = {A7},
        pages = {8045-8050},
          doi = {10.1029/JA091iA07p08045},
       adsurl = {https://ui.adsabs.harvard.edu/abs/1986JGR....91.8045L}
}

@article{Marsch1983,
       author = {{Marsch}, E. and {Goldstein}, H.},
        title = "{The effects of Coulomb collisions on solar wind ion velocity distributions}",
      journal = {\jgr},
         year = 1983,
        month = dec,
       volume = {88},
       number = {A12},
        pages = {9933-9940},
          doi = {10.1029/JA088iA12p09933},
       adsurl = {https://ui.adsabs.harvard.edu/abs/1983JGR....88.9933M}
}

@article{Hernandez1985,
       author = {{Hernandez}, R. and {Marsch}, E.},
        title = "{Collisional time scales for temperature and velocity exchange between drifting Maxwellians}",
      journal = {\jgr},
         year = 1985,
        month = nov,
       volume = {90},
       number = {A11},
        pages = {11062-11066},
          doi = {10.1029/JA090iA11p11062},
       adsurl = {https://ui.adsabs.harvard.edu/abs/1985JGR....9011062H}
}

@article{Klein1985,
       author = {{Klein}, L.~W. and {Ogilvie}, K.~W. and {Burlaga}, L.~F.},
        title = "{Coulomb collisions in the solar wind}",
      journal = {\jgr},
         year = 1985,
        month = aug,
       volume = {90},
       number = {A8},
        pages = {7389-7396},
          doi = {10.1029/JA090iA08p07389},
       adsurl = {https://ui.adsabs.harvard.edu/abs/1985JGR....90.7389K}
}

@article{Tracy2015,
       author = {{Tracy}, Patrick J. and {Kasper}, Justin C. and {Zurbuchen}, Thomas H. and {Raines}, Jim M. and {Shearer}, Paul and {Gilbert}, Jason},
        title = "{Thermalization of Heavy Ions in the Solar Wind}",
      journal = {\apj},
         year = 2015,
        month = oct,
       volume = {812},
       number = {2},
          eid = {170},
        pages = {170},
          doi = {10.1088/0004-637X/812/2/170},
       adsurl = {https://ui.adsabs.harvard.edu/abs/2015ApJ...812..170T}
}

@article{Alterman2018,
       author = {{Alterman}, B.~L. and {Kasper}, Justin C. and {Stevens}, Michael L. and {Koval}, Andriy},
        title = "{A Comparison of Alpha Particle and Proton Beam Differential Flows in Collisionally Young Solar Wind}",
      journal = {\apj},
         year = 2018,
        month = sep,
       volume = {864},
       number = {2},
          eid = {112},
        pages = {112},
          doi = {10.3847/1538-4357/aad23f},
archivePrefix = {arXiv},
       eprint = {1809.01693},
 primaryClass = {astro-ph.SR},
       adsurl = {https://ui.adsabs.harvard.edu/abs/2018ApJ...864..112A}
}

@article{Durovcova2021,
       author = {{{\v{D}}urovcov{\'a}}, Tereza and {{\v{S}}afr{\'a}nkov{\'a}}, Jana and {N{\v{e}}me{\v{c}}ek}, Zden{\v{e}}k},
        title = "{Proton Beam Abundance Variations and Their Relation to Alpha Particle Properties}",
      journal = {\apj},
         year = 2021,
        month = dec,
       volume = {923},
       number = {2},
          eid = {170},
        pages = {170},
          doi = {10.3847/1538-4357/ac2c03},
       adsurl = {https://ui.adsabs.harvard.edu/abs/2021ApJ...923..170D}
}

@article{mostafavi_et_al_2024,
       author = {{Mostafavi}, P. and {Allen}, R.~C. and {Jagarlamudi}, V.~K. and {Bourouaine}, S. and {Badman}, S.~T. and {Ho}, G.~C. and {Raouafi}, N.~E. and {Hill}, M.~E. and {Verniero}, J.~L. and {Larson}, D.~E. and et al.},
        title = "{Parker Solar Probe observations of collisional effects on thermalizing the young solar wind}",
      journal = {\aap},
         year = 2024,
        month = feb,
       volume = {682},
          eid = {A152},
        pages = {A152},
          doi = {10.1051/0004-6361/202347134},
       adsurl = {https://ui.adsabs.harvard.edu/abs/2024A&A...682A.152M}
}

@article{mario_et_al_2024,
       author = {{Amaro}, M{\'a}rio B. and {Vaivads}, Andris},
        title = "{Alpha-to-proton Temperature Ratio Distributions Using Parker Solar Probe Measurements}",
      journal = {\apjl},
         year = 2024,
        month = mar,
       volume = {964},
       number = {1},
          eid = {L2},
        pages = {L2},
          doi = {10.3847/2041-8213/ad2ded},
       adsurl = {https://ui.adsabs.harvard.edu/abs/2024ApJ...964L...2A}
}

@article{peng_et_al_2024,
       author = {{Peng}, Jingyu and {He}, Jiansen and {Duan}, Die and {Verscharen}, Daniel},
        title = "{Observations of Preferential Heating and Acceleration of {\ensuremath{\alpha}}-particles in the Young Solar Wind by Parker Solar Probe}",
      journal = {\apj},
         year = 2024,
        month = dec,
       volume = {977},
       number = {1},
          eid = {27},
        pages = {27},
          doi = {10.3847/1538-4357/ad79fa},
       adsurl = {https://ui.adsabs.harvard.edu/abs/2024ApJ...977...27P}
}

@article{Kasper2019,
       author = {{Kasper}, Justin C. and {Klein}, Kristopher G.},
        title = "{Strong Preferential Ion Heating is Limited to within the Solar Alfv{\'e}n Surface}",
      journal = {\apjl},
         year = 2019,
        month = jun,
       volume = {877},
       number = {2},
          eid = {L35},
        pages = {L35},
          doi = {10.3847/2041-8213/ab1de5},
archivePrefix = {arXiv},
       eprint = {1906.02763},
 primaryClass = {physics.space-ph},
       adsurl = {https://ui.adsabs.harvard.edu/abs/2019ApJ...877L..35K}
}

@article{Ogilvie1995,
       author = {{Ogilvie}, K.~W. and {Chornay}, D.~J. and {Fritzenreiter}, R.~J. and {Hunsaker}, F. and {Keller}, J. and {Lobell}, J. and {Miller}, G. and {Scudder}, J.~D. and {Sittler}, Jr., E.~C. and {Torbert}, R.~B. and et al.},
        title = "{SWE, A Comprehensive Plasma Instrument for the Wind Spacecraft}",
      journal = {\ssr},
         year = 1995,
        month = feb,
       volume = {71},
       number = {1-4},
        pages = {55-77},
          doi = {10.1007/BF00751326},
       adsurl = {https://ui.adsabs.harvard.edu/abs/1995SSRv...71...55O}
}

@article{Acuna1995,
  title={The global geospace science program and its investigations},
  author={Acu{\~n}a, MH and Ogilvie, KW and Baker, DN and Curtis, SA and Fairfield, DH and Mish, WH},
  journal={Space Science Reviews},
  volume={71},
  number={1},
  pages={5--21},
  year={1995},
  doi = {10.1007/BF00751323},
  publisher={Springer}
}

@misc{SILSO_Sunspot_Number,
  author = {{Clette}, F. and {Lefèvre}, L.},
  title = {SILSO Sunspot Number V2.0},
  howpublished = {https://doi.org/10.24414/qnza-ac80},
  month = {07},
  year = {2015},
  note = {Published by WDC SILSO - Royal Observatory of Belgium (ROB)}
}

@article{Kohl1998,
       author = {{Kohl}, J.~L. and {Noci}, G. and {Antonucci}, E. and {Tondello}, G. and {Huber}, M.~C.~E. and {Cranmer}, S.~R. and {Strachan}, L. and {Panasyuk}, A.~V. and {Gardner}, L.~D. and {Romoli}, M. and et al.},
        title = "{UVCS/SOHO Empirical Determinations of Anisotropic Velocity Distributions in the Solar Corona}",
      journal = {\apjl},
         year = 1998,
        month = jul,
       volume = {501},
       number = {1},
        pages = {L127-L131},
          doi = {10.1086/311434},
       adsurl = {https://ui.adsabs.harvard.edu/abs/1998ApJ...501L.127K}
}

@article{Ofman2010,
       author = {{Ofman}, Leon},
        title = "{Wave Modeling of the Solar Wind}",
      journal = {Living Reviews in Solar Physics},
         year = 2010,
        month = dec,
       volume = {7},
       number = {1},
          eid = {4},
        pages = {4},
          doi = {10.12942/lrsp-2010-4},
       adsurl = {https://ui.adsabs.harvard.edu/abs/2010LRSP....7....4O}
}

@article{Cranmer2012,
       author = {{Cranmer}, Steven R.},
        title = "{Self-Consistent Models of the Solar Wind}",
      journal = {\ssr},
         year = 2012,
        month = nov,
       volume = {172},
       number = {1-4},
        pages = {145-156},
          doi = {10.1007/s11214-010-9674-7},
archivePrefix = {arXiv},
       eprint = {1007.0954},
 primaryClass = {astro-ph.SR},
       adsurl = {https://ui.adsabs.harvard.edu/abs/2012SSRv..172..145C}
}

@ARTICLE{Maneva2013,
       author = {{Maneva}, Y.~G. and {Vi{\~n}As}, A.~F. and {Ofman}, L.},
        title = "{Turbulent heating and acceleration of He$^{++}$ ions by spectra of Alfv{\'e}n-cyclotron waves in the expanding solar wind: 1.5-D hybrid simulations}",
      journal = {Journal of Geophysical Research (Space Physics)},
         year = 2013,
        month = jun,
       volume = {118},
       number = {6},
        pages = {2842-2853},
          doi = {10.1002/jgra.50363},
       adsurl = {https://ui.adsabs.harvard.edu/abs/2013JGRA..118.2842M}
}

@ARTICLE{Abbo2016,
       author = {{Abbo}, L. and {Ofman}, L. and {Antiochos}, S.~K. and {Hansteen}, V.~H. and {Harra}, L. and {Ko}, Y.-K. and {Lapenta}, G. and {Li}, B. and {Riley}, P. and {Strachan}, L. and {von Steiger}, R. and {Wang}, Y.-M.},
        title = "{Slow Solar Wind: Observations and Modeling}",
      journal = {\ssr},
         year = 2016,
        month = nov,
       volume = {201},
       number = {1-4},
        pages = {55-108},
          doi = {10.1007/s11214-016-0264-1},
       adsurl = {https://ui.adsabs.harvard.edu/abs/2016SSRv..201...55A}
}

@article{Antiochos2011,
       author = {{Antiochos}, S.~K. and {Miki{\'c}}, Z. and {Titov}, V.~S. and {Lionello}, R. and {Linker}, J.~A.},
        title = "{A Model for the Sources of the Slow Solar Wind}",
      journal = {\apj},
         year = 2011,
        month = apr,
       volume = {731},
       number = {2},
          eid = {112},
        pages = {112},
          doi = {10.1088/0004-637X/731/2/112},
archivePrefix = {arXiv},
       eprint = {1102.3704},
 primaryClass = {astro-ph.SR},
       adsurl = {https://ui.adsabs.harvard.edu/abs/2011ApJ...731..112A}
}

@article{McComas2008,
       author = {{McComas}, D.~J. and {Ebert}, R.~W. and {Elliott}, H.~A. and {Goldstein}, B.~E. and {Gosling}, J.~T. and {Schwadron}, N.~A. and {Skoug}, R.~M.},
        title = "{Weaker solar wind from the polar coronal holes and the whole Sun}",
      journal = {\grl},
         year = 2008,
        month = sep,
       volume = {35},
       number = {18},
          eid = {L18103},
        pages = {L18103},
          doi = {10.1029/2008GL034896},
       adsurl = {https://ui.adsabs.harvard.edu/abs/2008GeoRL..3518103M}
}

@article{Wang2009,
       author = {{Wang}, Y.-M. and {Ko}, Y.-K. and {Grappin}, R.},
        title = "{Slow Solar Wind from Open Regions with Strong Low-Coronal Heating}",
      journal = {\apj},
         year = 2009,
        month = jan,
       volume = {691},
       number = {1},
        pages = {760-769},
          doi = {10.1088/0004-637X/691/1/760},
       adsurl = {https://ui.adsabs.harvard.edu/abs/2009ApJ...691..760W}
}

@article{McComas2003,
       author = {{McComas}, D.~J. and {Elliott}, H.~A. and {Schwadron}, N.~A. and {Gosling}, J.~T. and {Skoug}, R.~M. and {Goldstein}, B.~E.},
        title = "{The three-dimensional solar wind around solar maximum}",
      journal = {\grl},
         year = 2003,
        month = may,
       volume = {30},
       number = {10},
          eid = {1517},
        pages = {1517},
          doi = {10.1029/2003GL017136},
       adsurl = {https://ui.adsabs.harvard.edu/abs/2003GeoRL..30.1517M}
}

@article{Alterman2019,
       author = {{Alterman}, B.~L. and {Kasper}, Justin C.},
        title = "{Helium Variation across Two Solar Cycles Reveals a Speed-dependent Phase Lag}",
      journal = {\apjl},
         year = 2019,
        month = jul,
       volume = {879},
       number = {1},
          eid = {L6},
        pages = {L6},
          doi = {10.3847/2041-8213/ab2391},
archivePrefix = {arXiv},
       eprint = {1906.12273},
 primaryClass = {physics.space-ph},
       adsurl = {https://ui.adsabs.harvard.edu/abs/2019ApJ...879L...6A}
}

@ARTICLE{Fu2018,
       author = {{Fu}, Hui and {Madjarska}, Maria S. and {Li}, Bo and {Xia}, Lidong and {Huang}, Zhenghua},
        title = "{Helium abundance and speed difference between helium ions and protons in the solar wind from coronal holes, active regions, and quiet Sun}",
      journal = {\mnras},
         year = 2018,
        month = aug,
       volume = {478},
       number = {2},
        pages = {1884-1892},
          doi = {10.1093/mnras/sty1211},
archivePrefix = {arXiv},
       eprint = {1805.02880},
 primaryClass = {astro-ph.SR},
       adsurl = {https://ui.adsabs.harvard.edu/abs/2018MNRAS.478.1884F}
}

@ARTICLE{Jagarlamudi2025,
       author = {{Jagarlamudi}, Vamsee Krishna and {Mostafavi}, P. and {Palacios}, J.~C. and {Bourouaine}, S. and {Raouafi}, N.~E. and {Kieokaew}, Rungployphan and {{\v{D}}urovcov{\'a}}, Tereza and {Fedorov}, Andrey and {Louarn}, Philippe},
        title = "{Probing the Evolution of Solar Wind Temperature Anisotropies in the Inner Heliosphere}",
      journal = {\apjl},
         year = 2025,
        month = dec,
       volume = {995},
       number = {2},
          eid = {L68},
        pages = {L68},
          doi = {10.3847/2041-8213/ae2792},
       adsurl = {https://ui.adsabs.harvard.edu/abs/2025ApJ...995L..68J}
}

@article{Mostafavi2025,
       author = {{Mostafavi}, Parisa and {Jagarlamudi}, V.~K. and {Raouafi}, N.~E. and {Palacios}, J.~C. and {Allen}, R.~C. and {Hill}, M.~E. and {Ofman}, L. and {Ho}, G.~C.},
        title = "{Preferential Energization of Solar Wind Ions Below the Alfv{\'e}nic Surface}",
      journal = {\apjl},
         year = 2025,
        month = oct,
       volume = {991},
       number = {2},
          eid = {L35},
        pages = {L35},
          doi = {10.3847/2041-8213/ae0732},
       adsurl = {https://ui.adsabs.harvard.edu/abs/2025ApJ...991L..35M}
}

@article{Mihailo2026,
  title = {How the Oblique Drift Instability Alters Solar Wind Heating and Constrains the Distribution of Solar Wind Observations},
  author = {Martinovi\ifmmode \acute{c}\else \'{c}\fi{}, Mihailo M. and Klein, Kristopher G. and Ofman, Leon and Yogesh and Howes, Gregory G. and Verniero, Jaye L. and Yoon, Peter H. and Verscharen, Daniel and Alterman, Benjamin L.},
  journal = {Phys. Rev. Lett.},
  volume = {136},
  issue = {25},
  pages = {255201},
  numpages = {8},
  year = {2026},
  month = {Jun},
  publisher = {American Physical Society},
  doi = {10.1103/nlq9-2g37},
  url = {https://link.aps.org/doi/10.1103/nlq9-2g37}
}

@ARTICLE{Alterman2025,
       author = {{Alterman}, B.~L. and {Rivera}, Y.~J. and {Lepri}, S.~T. and {Raines}, J.~M.},
        title = "{The transition from slow to fast wind as observed in composition observations}",
      journal = {\aap},
         year = 2025,
        month = feb,
       volume = {694},
          eid = {A265},
        pages = {A265},
          doi = {10.1051/0004-6361/202451550},
archivePrefix = {arXiv},
       eprint = {2411.18984},
 primaryClass = {astro-ph.SR},
       adsurl = {https://ui.adsabs.harvard.edu/abs/2025A&A...694A.265A}
}

@ARTICLE{Alterman2021,
       author = {{Alterman}, Benjamin L. and {Kasper}, Justin C. and {Leamon}, Robert J. and {McIntosh}, Scott W.},
        title = "{Solar Wind Helium Abundance Heralds Solar Cycle Onset}",
      journal = {\solphys},
         year = 2021,
        month = apr,
       volume = {296},
       number = {4},
          eid = {67},
        pages = {67},
          doi = {10.1007/s11207-021-01801-9},
archivePrefix = {arXiv},
       eprint = {2006.04669},
 primaryClass = {astro-ph.SR},
       adsurl = {https://ui.adsabs.harvard.edu/abs/2021SoPh..296...67A}
}

@ARTICLE{Aellig2001,
       author = {{Aellig}, Matthias R. and {Lazarus}, Alan J. and {Steinberg}, John T.},
        title = "{The solar wind helium abundance: Variation with wind speed and the solar cycle}",
      journal = {\grl},
         year = 2001,
        month = jul,
       volume = {28},
       number = {14},
        pages = {2767-2770},
          doi = {10.1029/2000GL012771},
       adsurl = {https://ui.adsabs.harvard.edu/abs/2001GeoRL..28.2767A}
}

@ARTICLE{Kasper2012,
       author = {{Kasper}, J.~C. and {Stevens}, M.~L. and {Korreck}, K.~E. and {Maruca}, B.~A. and {Kiefer}, K.~K. and {Schwadron}, N.~A. and {Lepri}, S.~T.},
        title = "{Evolution of the Relationships between Helium Abundance, Minor Ion Charge State, and Solar Wind Speed over the Solar Cycle}",
      journal = {\apj},
         year = 2012,
        month = feb,
       volume = {745},
       number = {2},
          eid = {162},
        pages = {162},
          doi = {10.1088/0004-637X/745/2/162},
       adsurl = {https://ui.adsabs.harvard.edu/abs/2012ApJ...745..162K}
}

@ARTICLE{Kasper2007,
       author = {{Kasper}, Justin C. and {Stevens}, Michael L. and {Lazarus}, Alan J. and {Steinberg}, John T. and {Ogilvie}, Keith. W.},
        title = "{Solar Wind Helium Abundance as a Function of Speed and Heliographic Latitude: Variation through a Solar Cycle}",
      journal = {\apj},
         year = 2007,
        month = may,
       volume = {660},
       number = {1},
        pages = {901-910},
          doi = {10.1086/510842},
       adsurl = {https://ui.adsabs.harvard.edu/abs/2007ApJ...660..901K}
}

@ARTICLE{Zerbo2015,
       author = {{Zerbo}, J.-L. and {Richardson}, J.~D.},
        title = "{The solar wind during current and past solar minima and maxima}",
      journal = {Journal of Geophysical Research (Space Physics)},
         year = 2015,
        month = dec,
       volume = {120},
       number = {12},
        pages = {10,250-10,256},
          doi = {10.1002/2015JA021407},
       adsurl = {https://ui.adsabs.harvard.edu/abs/2015JGRA..12010250Z}
}

@ARTICLE{Johnson2024,
       author = {{Johnson}, E. and {Maruca}, B.~A. and {McManus}, M. and {Stevens}, M. and {Klein}, K.~G. and {Mostafavi}, P.},
        title = "{Application of collisional analysis to the differential velocity of solar wind ions}",
      journal = {Frontiers in Astronomy and Space Sciences},
         year = 2024,
        month = jan,
       volume = {10},
          eid = {1284913},
        pages = {1284913},
          doi = {10.3389/fspas.2023.1284913},
       adsurl = {https://ui.adsabs.harvard.edu/abs/2024FrASS..1084913J}
}

@ARTICLE{Johnson2023,
       author = {{Johnson}, E. and {Maruca}, B.~A. and {McManus}, M. and {Klein}, K.~G. and {Lichko}, E.~R. and {Verniero}, J. and {Paulson}, K.~W. and {DeWeese}, H. and {Dieguez}, I. and {Qudsi}, R.~A. and et al.},
        title = "{Anterograde Collisional Analysis of Solar Wind Ions}",
      journal = {\apj},
         year = 2023,
        month = jun,
       volume = {950},
       number = {1},
          eid = {51},
        pages = {51},
          doi = {10.3847/1538-4357/accc32},
       adsurl = {https://ui.adsabs.harvard.edu/abs/2023ApJ...950...51J}
}

@ARTICLE{Rivera2024,
       author = {{Rivera}, Yeimy J. and {Badman}, Samuel T. and {Stevens}, Michael L. and {Verniero}, Jaye L. and {Stawarz}, Julia E. and {Shi}, Chen and {Raines}, Jim M. and {Paulson}, Kristoff W. and {Owen}, Christopher J. and {Niembro}, Tatiana and et al.},
        title = "{In situ observations of large-amplitude Alfv{\'e}n waves heating and accelerating the solar wind}",
      journal = {Science},
         year = 2024,
        month = aug,
       volume = {385},
       number = {6712},
        pages = {962-966},
          doi = {10.1126/science.adk6953},
archivePrefix = {arXiv},
       eprint = {2409.00267},
 primaryClass = {astro-ph.SR},
       adsurl = {https://ui.adsabs.harvard.edu/abs/2024Sci...385..962R}
}

@ARTICLE{Yardley2024,
       author = {{Yardley}, Stephanie L. and {Brooks}, David H. and {D'Amicis}, Raffaella and {Owen}, Christopher J. and {Long}, David M. and {Baker}, Deb and {D{\'e}moulin}, Pascal and {Owens}, Mathew J. and {Lockwood}, Mike and {Mihailescu}, Teodora and et al.},
        title = "{Multi-source connectivity as the driver of solar wind variability in the heliosphere}",
      journal = {Nature Astronomy},
         year = 2024,
        month = aug,
       volume = {8},
        pages = {953-963},
          doi = {10.1038/s41550-024-02278-9},
       adsurl = {https://ui.adsabs.harvard.edu/abs/2024NatAs...8..953Y}
}

@article{Rivera2025a,
  title={Differentiating the acceleration mechanisms in the slow and Alfv{\'e}nic slow solar wind},
  author={Rivera, Yeimy J and Badman, Samuel T and Verniero, JL and Varesano, Tania and Stevens, Michael L and Stawarz, Julia E and Reeves, Katharine K and Raines, Jim M and Raymond, John C and Owen, Christopher J and others},
  journal={The Astrophysical Journal},
  volume={980},
  number={1},
  pages={70},
  year={2025},
  doi = {10.3847/1538-4357/ada699},
  publisher={The American Astronomical Society}
}

@article{Rivera2025b,
  title={Observational Constraints on the Radial Evolution of O6+ Temperature and Differential Flow in the Inner Heliosphere},
  author={Rivera, Yeimy J and Klein, Kristopher G and Wang, Joseph H and Matteini, Lorenzo and Verscharen, Daniel and Coburn, Jesse T and Badman, Samuel T and Lepri, Susan T and Dewey, Ryan M and Raines, Jim M and others},
  journal={The Astrophysical Journal Letters},
  volume={990},
  number={2},
  pages={L60},
  year={2025},
  doi = {10.3847/2041-8213/adfa97},
  publisher={The American Astronomical Society}
}

@article{Alterman2025a,
  title={Heavy ion abundances evolve with solar activity},
  author={Alterman, BL and Rivera, YJ and Lepri, ST and Raines, JM and D’Amicis, R},
  journal={Astronomy \& Astrophysics},
  volume={700},
  pages={A23},
  year={2025},
  doi = {10.1051/0004-6361/202554299},
  publisher={EDP Sciences}
}

@article{Alterman2025b,
  title={Cross Helicity and the Helium Abundance as an In Situ Metric of Solar Wind Acceleration},
  author={Alterman, BL and D’Amicis, Raffaella},
  journal={The Astrophysical Journal Letters},
  volume={982},
  number={2},
  pages={L40},
  year={2025},
  doi = {10.3847/2041-8213/adb48e},
  publisher={The American Astronomical Society}
}

@article{Alterman2026,
  title={On the Regulation of the Solar Wind Helium Abundance by the Hydrogen Compressibility},
  author={Alterman, BL and D’Amicis, Raffaella},
  journal={The Astrophysical Journal Letters},
  volume={996},
  number={1},
  pages={L12},
  year={2026},
  doi = {10.3847/2041-8213/ae002f},
  publisher={The American Astronomical Society}
}

@article{Ben2019,
  title={The significance of proton beams in the multiscale solar wind},
  author={Alterman, Benjamin L},
  journal={Ph. D. Thesis},
  year={2019}
}

@article{Heidrich2020,
  title={Proton-proton collisional age to order solar wind types},
  author={Heidrich-Meisner, Verena and Berger, Lars and Wimmer-Schweingruber, Robert F},
  journal={Astronomy \& Astrophysics},
  volume={636},
  pages={A103},
  year={2020},
  doi = {10.1051/0004-6361/201937378},
  publisher={EDP Sciences}
}

@article{Xu2015,
  title={A new four-plasma categorization scheme for the solar wind},
  author={Xu, Fei and Borovsky, Joseph E},
  journal={Journal of Geophysical Research: Space Physics},
  volume={120},
  number={1},
  pages={70--100},
  year={2015},
  doi = {10.1002/2014JA020412},
  publisher={Wiley Online Library}
}

@ARTICLE{Martinovic2020,
       author = {{Martinovi{\'c}}, Mihailo M. and {Klein}, Kristopher G. and {Kasper}, Justin C. and {Case}, Anthony W. and {Korreck}, Kelly E. and {Larson}, Davin and {Livi}, Roberto and {Stevens}, Michael and {Whittlesey}, Phyllis and {Chandran}, Benjamin D.~G. and et al.},
        title = "{The Enhancement of Proton Stochastic Heating in the Near-Sun Solar Wind}",
      journal = {\apjs},
         year = 2020,
        month = feb,
       volume = {246},
       number = {2},
          eid = {30},
        pages = {30},
          doi = {10.3847/1538-4365/ab527f},
archivePrefix = {arXiv},
       eprint = {1912.02653},
 primaryClass = {astro-ph.SR},
       adsurl = {https://ui.adsabs.harvard.edu/abs/2020ApJS..246...30M}
}

@ARTICLE{Bowen2025,
       author = {{Bowen}, Trevor A. and {Ervin}, Tamar and {Mallet}, Alfred and {Chandran}, Benjamin D.~G. and {Sioulas}, Nikos and {Isenberg}, Philip A. and {Bale}, Stuart D. and {Squire}, Jonathan and {Klein}, Kristopher G. and {Pezzi}, Oreste},
        title = "{Stochastic Heating in the Sub-Alfv{\'e}nic Solar Wind}",
      journal = {\prl},
         year = 2025,
        month = dec,
       volume = {135},
       number = {25},
          eid = {255201},
        pages = {255201},
          doi = {10.1103/rxd8-22m9},
archivePrefix = {arXiv},
       eprint = {2509.20654},
 primaryClass = {astro-ph.SR},
       adsurl = {https://ui.adsabs.harvard.edu/abs/2025PhRvL.135y5201B}
}

@ARTICLE{Martinovic2025,
       author = {{Martinovic}, Mihailo M. and {Klein}, Kristopher G. and {Ofman}, Leon and {Yogesh} and {Verniero}, Jaye L. and {Yoon}, Peter H. and {Howes}, Gregory G. and {Verscharen}, Daniel and {Alterman}, Benjamin L.},
        title = "{How the Oblique Drift Instability Alters Solar Wind Heating and Constrains the Distribution of Solar Wind Observations}",
      journal = {arXiv e-prints},
         year = 2025,
        month = dec,
          eid = {arXiv:2512.18485},
        pages = {arXiv:2512.18485},
          doi = {10.48550/arXiv.2512.18485},
archivePrefix = {arXiv},
       eprint = {2512.18485},
 primaryClass = {astro-ph.SR},
       adsurl = {https://ui.adsabs.harvard.edu/abs/2025arXiv251218485M}
}

@article{Mostafavi2024,
  title={Parker Solar Probe observations of collisional effects on thermalizing the young solar wind},
  author={Mostafavi, P and Allen, RC and Jagarlamudi, VK and Bourouaine, S and Badman, ST and Ho, GC and Raouafi, NE and Hill, ME and Verniero, JL and Larson, DE and others},
  journal={Astronomy \& Astrophysics},
  volume={682},
  pages={A152},
  year={2024},
  doi = {10.1051/0004-6361/202347134},
  publisher={EDP Sciences}
}

@article{Johnson2025,
  title={Collisional thermalization of minor ions in the solar wind},
  author={Johnson, Elliot and Maruca, BA},
  journal={Frontiers in Astronomy and Space Sciences},
  volume={12},
  pages={1586421},
  year={2025},
  publisher={Frontiers Media SA}
}

@article{DAmicis2021,
  title={First Solar Orbiter observation of the Alfv{\'e}nic slow wind and identification of its solar source},
  author={d’Amicis, R and Bruno, R and Panasenco, O and Telloni, D and Perrone, D and Marcucci, MF and Woodham, L and Velli, M and De Marco, R and Jagarlamudi, V and others},
  journal={Astronomy \& Astrophysics},
  volume={656},
  pages={A21},
  year={2021},
  publisher={EDP Sciences}
}

@article{Ogilvie1974,
  title={The solar cycle variation of the solar wind helium abundance},
  author={Ogilvie, KW and Hirshberg, J},
  journal={Journal of Geophysical Research},
  volume={79},
  number={31},
  pages={4595--4602},
  year={1974},
  publisher={Wiley Online Library}
}

@article{Yogesh2026,
  title={Solar Wind Heating near the Sun: A Radial Evolution Approach},
  author={Yogesh and Ofman, Leon and Klein, Kristopher G and Niranjana and Martinovi{\'c}, Mihailo and Howes, Gregory G and Mostafavi, Parisa and Boardsen, Scott A and Sadykov, Viacheslav M and Pal, Sanchita and others},
  journal={The Astrophysical Journal},
  volume={999},
  number={2},
  pages={225},
  year={2026},
  doi = {10.3847/1538-4357/ae4582},
  publisher={The American Astronomical Society}
}

@article{Sebastian2025,
  title={Comparison of Solar Wind flux and bulk parameters obtained from Aditya L1-ASPEX with Wind-3DP PESA-L and ACE-EPAM-LEMS120},
   author={Sebastian, Jacob and Kumar, Abhishek and Chakrabarty, Dibyendu and Kumar, Prashant and Goyal, Shiv Kumar and Vadawale, Santosh and Shanmugam, M. and Hasan, M. and Sarkar, Aveek and Dalal, Bijoy and Gupta, Aakash and Parashar, Shivam and Bapat, Bhas and Shah, Manan S. and Adhyaru, Pranav R. and Patel, Arpit R. and Adalja, Hitesh Kumar and Tiwari, Neeraj Kumar and Sarda, Aaditya and Sharma, Piyush and Ladiya, Tinkal and Kumar, Sushil and Singh, Nishant and Painkra, Deepak Kumar and Verma, Abhishek J. and Banerjee, Swaroop and Subramanian, K. P. and Dadhania, M. B. and Janardhan, P. and Bhardwaj, Anil},
  journal      = {GSICS Quarterly},
  volume       = {18},
  number       = {4},
  year         = {2025},
  publisher    = {Global Space-Based Inter-Calibration System Coordination Center and Center for Satellite Applications and Research (U.S.)},
  doi          = {10.25923/gmzc-9a28}
}

@article{Sebastian2026,
  title={One Year of ASPEX-STEPS Operation: Characteristic Features, Observations and Science Potential},
  author={Sebastian, Jacob and Dalal, Bijoy and Gupta, Aakash and Goyal, Shiv Kumar and Chakrabarty, Dibyendu and Vadawale, Santosh V and Shanmugam, M and Tiwari, Neeraj Kumar and Patel, Arpit R and Sarkar, Aveek and others},
  journal={Journal of Astrophysics and Astronomy},
  volume={47},
  number={1},
  pages={21},
  year={2026},
  doi = {10.1007/s12036-026-10134-7},
  publisher={Springer}
}

@article{Gupta2025,
  title={Multidirectional Investigations on Quiet Time Suprathermal Ions Measured by ASPEX-STEPS on Board Aditya L1},
  author={Gupta, Aakash and Chakrabarty, Dibyendu and Vadawale, Santosh and Sarkar, Aveek and Dalal, Bijoy and Goyal, Shiv Kumar and Sebastian, Jacob and Janardhan, P and Srivastava, Nandita and Shanmugam, M and others},
  journal={The Astrophysical Journal},
  volume={995},
  number={1},
  pages={92},
  year={2025},
  doi = {10.3847/1538-4357/ae1a46},
  publisher={The American Astronomical Society}
}

@article{Parashar2026,
  title={Evidence for In Situ Particle Energization during the 2024 May Event Based on the ASPEX Instrument on Board Aditya-L1},
  author={Parashar, Shivam and Chakrabarty, Dibyendu and Kumar, Prashant and Kumar, Abhishek and Bapat, Bhas and Sarkar, Aveek and Janardhan, P and Bhardwaj, Anil and Vadawale, Santosh V and Shah, Manan S and others},
  journal={The Astrophysical Journal Letters},
  volume={996},
  number={2},
  pages={L36},
  year={2026},
  doi = {10.3847/2041-8213/ae2ac9},
  publisher={The American Astronomical Society}
}

@ARTICLE{Kumar2025,
       author = {{Kumar}, Prashant and {Bapat}, Bhas and {Shah}, Manan S. and {Adalja}, Hiteshkumar L. and {Patel}, Arpit R. and {Adhyaru}, Pranav R. and {Shanmugam}, M. and {Chakrabarty}, Dibyendu and {Banerjee}, Swaroop B. and {Subramanian}, K.~P. and et al.},
        title = "{Aditya Solar Wind Particle Experiment (ASPEX) on Board Aditya{\textemdash}L1: The Solar Wind Ion Spectrometer (SWIS)}",
      journal = {\solphys},
         year = 2025,
        month = apr,
       volume = {300},
       number = {4},
          eid = {37},
        pages = {37},
          doi = {10.1007/s11207-025-02443-x},
       adsurl = {https://ui.adsabs.harvard.edu/abs/2025SoPh..300...37K}
}

@ARTICLE{Goyal2025,
       author = {{Goyal}, Shiv Kumar and {Tiwari}, Neeraj Kumar and {Patel}, Arpit R. and {Shanmugam}, M. and {Vadawale}, Santosh V. and {Chakrabarty}, Dibyendu and {Sebastian}, Jacob and {Dalal}, Bijoy and {Sharma}, Piyush and {Sarkar}, Aveek and et al.},
        title = "{Aditya Solar Wind Particle Experiment on Board Aditya─L1: The Supra-Thermal and Energetic Particle Spectrometer}",
      journal = {\solphys},
         year = 2025,
        month = mar,
       volume = {300},
       number = {3},
          eid = {35},
        pages = {35},
          doi = {10.1007/s11207-025-02441-z},
       adsurl = {https://ui.adsabs.harvard.edu/abs/2025SoPh..300...35G}
}
\bibliographystyle{aasjournalv7}

\end{document}